\documentclass[aps,prl,twocolumn,superscriptaddress]{revtex4-1}

\long\def\symbolfootnote[#1]#2{\begingroup%
\def\thefootnote{\fnsymbol{footnote}}\footnotetext[#1]{#2}\endgroup}

\usepackage[utf8x]{inputenc}
\usepackage{color}
\usepackage{bbm} 

\usepackage{amsfonts,amsmath,amssymb,stmaryrd}

\usepackage{graphicx}
\usepackage{subfigure}  
\usepackage{bbm} 
\usepackage{hyperref}
\usepackage{epsfig}
\usepackage{mathrsfs}
\usepackage{verbatim}
\usepackage{centernot}
\usepackage{ulem}
\usepackage{longtable}
\usepackage{nicefrac}
\usepackage[framemethod=tikz]{mdframed}

\usepackage{amsmath}
\usepackage{amssymb}
\usepackage{wasysym}
\usepackage{graphicx}
\usepackage{color}
\usepackage{physics}
\usepackage{siunitx}
\usepackage[english,nomargin,inline,marginclue,draft]{fixme}
\fxusetheme{colorsig}
\FXRegisterAuthor{cg}{acg}{CG}  
\makeatletter
\renewcommand*\FXLayoutInline[3]{%
  {\@fxuseface{inline}\ignorespaces{\color{fx#1}[#3: #2]}}}
\makeatother

\renewcommand{\l}{\left(}
\renewcommand{\r}{\right)}

\usepackage{array}

\usepackage{cancel,ifthen}
\newcommand{\cmnt}[2][NoInPuT]{\ifthenelse{\equal{#1}{NoInPuT}}{}{{\color{red}\sout{#1}}} {\color{blue} #2}}

\usepackage{bm}	
\renewcommand{\vec}[1]{\bm{#1}}

\newcommand{\Ham} {{\mathcal{H}}}

\begin{document}

\title{Microscopic evolution of doped Mott insulators \\
from polaronic metal to Fermi liquid }

\author{Joannis Koepsell}
\email{joannis.koepsell@mpq.mpg.de}
\affiliation{Max-Planck-Institut f\"{u}r Quantenoptik, 85748 Garching, Germany}
\affiliation{Munich Center for Quantum Science and Technology (MCQST), 80799 M\"{u}nchen, Germany}
\author{Dominik Bourgund}
\affiliation{Max-Planck-Institut f\"{u}r Quantenoptik, 85748 Garching, Germany}
\affiliation{Munich Center for Quantum Science and Technology (MCQST), 80799 M\"{u}nchen, Germany}
\author{Pimonpan Sompet}
\affiliation{Max-Planck-Institut f\"{u}r Quantenoptik, 85748 Garching, Germany}
\affiliation{Munich Center for Quantum Science and Technology (MCQST), 80799 M\"{u}nchen, Germany}
\author{Sarah Hirthe}
\affiliation{Max-Planck-Institut f\"{u}r Quantenoptik, 85748 Garching, Germany}
\affiliation{Munich Center for Quantum Science and Technology (MCQST), 80799 M\"{u}nchen, Germany}
\author{Annabelle Bohrdt}
\affiliation{Munich Center for Quantum Science and Technology (MCQST), 80799 M\"{u}nchen, Germany}
\affiliation{Department of Physics and Institute for Advanced Study,
Technical University of Munich, 85748 Garching, Germany}
\author{Yao Wang}
\affiliation{Department of Physics, Harvard University, Cambridge, Massachusetts 02138, USA}
\affiliation{Department of Physics and Astronomy, Clemson University, Clemson, South Carolina 29631, USA}
\author{Fabian Grusdt}
\affiliation{Munich Center for Quantum Science and Technology (MCQST), 80799 M\"{u}nchen, Germany}
\affiliation{Fakult\"{a}t f\"{u}r Physik, Ludwig-Maximilians-Universit\"{a}t, 80799 M\"{u}nchen, Germany}
\author{Eugene Demler}
\affiliation{Department of Physics, Harvard University, Cambridge, Massachusetts 02138, USA}
\author{Guillaume Salomon}
\affiliation{Max-Planck-Institut f\"{u}r Quantenoptik, 85748 Garching, Germany}
\affiliation{Munich Center for Quantum Science and Technology (MCQST), 80799 M\"{u}nchen, Germany}
\affiliation{Institut f\"{u}r Laserphysik, Universit\"{a}t Hamburg, Germany}
\affiliation{The Hamburg Centre for Ultrafast Imaging, Universit\"{a}t Hamburg, Luruper Chaussee 149, 22761 Hamburg, Germany}
\author{Christian Gross}
\affiliation{Max-Planck-Institut f\"{u}r Quantenoptik, 85748 Garching, Germany}
\affiliation{Munich Center for Quantum Science and Technology (MCQST), 80799 M\"{u}nchen, Germany}
\affiliation{Physikalisches Institut, Eberhard Karls Universit\"{a}t T\"{u}bingen, 72076 T\"{u}bingen, Germany}
\author{Immanuel Bloch}
\affiliation{Max-Planck-Institut f\"{u}r Quantenoptik, 85748 Garching, Germany}
\affiliation{Munich Center for Quantum Science and Technology (MCQST), 80799 M\"{u}nchen, Germany}
\affiliation{Fakult\"{a}t f\"{u}r Physik, Ludwig-Maximilians-Universit\"{a}t, 80799 M\"{u}nchen, Germany}


\begin{abstract}
The competition between antiferromagnetism and hole motion in two-dimensional Mott insulators lies at the heart of a doping-dependent transition from an anomalous metal to a conventional Fermi liquid. Condensed matter experiments suggest charge carriers change their nature within this crossover, but a complete understanding remains elusive. We observe such a crossover in Fermi-Hubbard systems on a cold-atom quantum simulator and reveal the transformation of multi-point correlations between spins and holes upon increasing doping at temperatures around the superexchange energy. Conventional observables, such as spin susceptibility, are furthermore computed from the microscopic snapshots of the system. Starting from a magnetic polaron regime, we find the system evolves into a Fermi liquid featuring incommensurate magnetic fluctuations and fundamentally altered correlations. The crossover is completed for hole dopings around $30\%$. Our work benchmarks theoretical approaches and discusses possible connections to lower temperature phenomena.
\end{abstract}

\maketitle
\begin{figure*}[]
\includegraphics{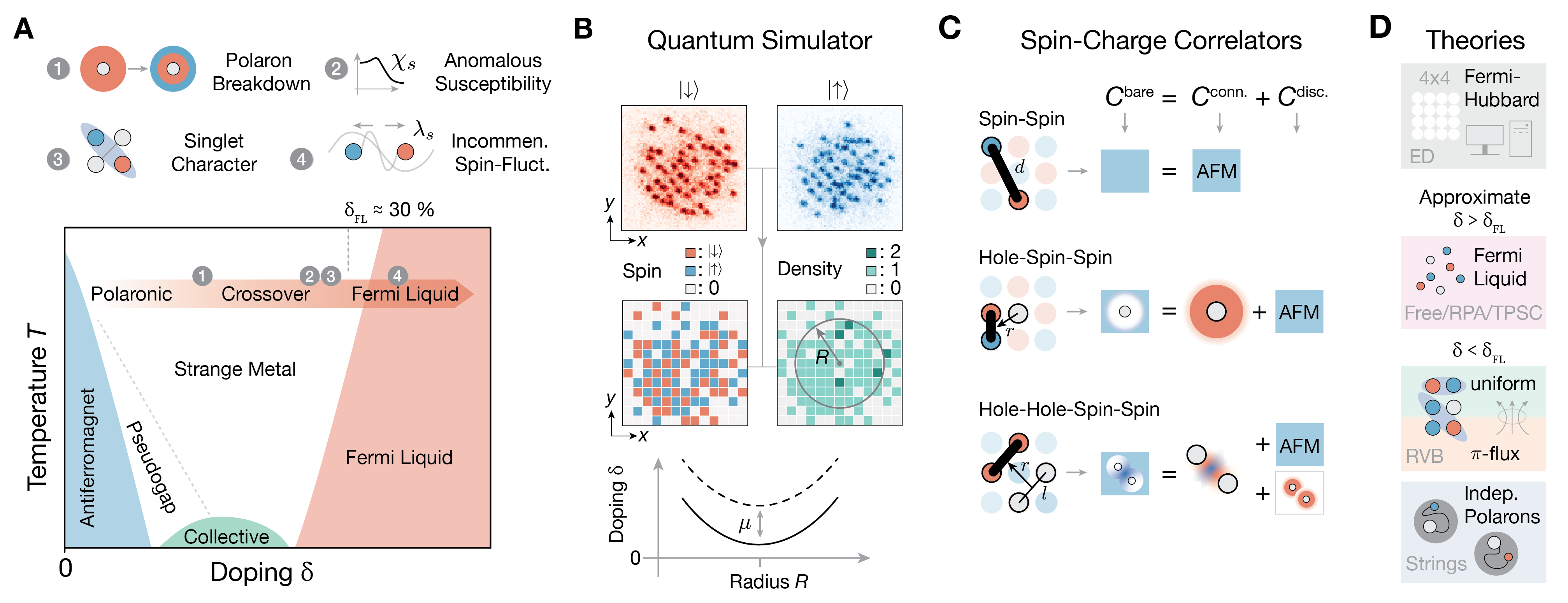}
\caption{\textbf{Probing doped Mott insulators with spin-charge correlators.} \textbf{A}, Conjectured phase diagram of the two-dimensional Fermi-Hubbard model upon hole doping $\delta$ and temperature $T$. Boundaries indicate crossovers between different regimes. Insets summarize our main results, which we obtain from, \textbf{B}, our quantum gas microscope with full spin and density (charge) resolution. We use the spatially varying doping in our harmonic trap and the control of total particle number to study the doping dependence of, \textbf{C}, connected spin-spin and spin-charge correlators consisting of up to four points. As illustrated, bare multi-point correlations are composed out of lower-order contributions and the connected correlation, which carries the new higher-order correlation information. \textbf{D}, We compare experimental findings to exact diagonalization of $4\times4$ Fermi-Hubbard systems, mean-field inspired approaches or free fermions approximating Fermi liquids at high doping as well as three approaches (uniform-RVB, $\pi$-flux and string), which are designed to capture the low doping regime.}
\label{fig:fig1}
\end{figure*}

Interacting electrons in conventional metals are successfully described by Landau's Fermi-liquid (FL) theory, which captures the universal behavior of macroscopic properties. The violation of these concepts is a hallmark of strongly-correlated quantum materials, leading to the appearance of pseudogap or strange metal regimes \cite{Keimer2015}.

Particularly interesting materials are doped antiferromagnetic Mott insulators, because they exhibit non-FL behavior for weak doping, but turn into normal FLs for high doping \cite{Dagotto1994, Keimer2015, Lee2006}. Furthermore, these systems often host unconventional superconductivity. The highest transition temperatures in hole-doped cuprates exists in the strange metal phase, which indicates a strong relation between the two phenomena.

Recent studies on cuprates suggest, that a transition from unconventional metal to FL occurs at a hole doping of $\delta^{\star}\approx20\, \%$ \cite{Badoux2016, Chen2019}, which is expected to be material dependent. Spectroscopy and transport measurements hint at charge carriers being `hole-like' below and `particle (electron)-like' above this hole concentration \cite{Doiron2007, Yang2009, Chen2019}. Nonetheless, the interpretation and universality of such findings is unclear, due to the microscopic complexity of real materials. 

In Mott insulators slightly below half filling, the competition between hole motion and antiferromagnetism leads to heavily dressed dopants \cite{Ronning2005, Schrieffer2007}, referred to as magnetic polarons \cite{Bulaevski1968,SchmittRink1988, Shraiman1988, Kane1989, Sachdev1989, Grusdt2018, Blomquist2019}. The interplay between magnetism and hole hopping prevails up to intermediate dopings \cite{Frachet2020} and is believed to ultimately trigger pseudogap and superconducting phases at colder temperatures \cite{Lee2006, Keimer2015}. A generally accepted description of these phenomena in terms of interacting magnetic polarons, spin-liquid states or other microscopic models remains elusive. For large dopings antiferromagnetic correlations become strongly suppressed, particle motion is restored in the dilute system and FL type quasiparticles form. At which hole concentration magnetic polarons dissolve, whether exotic regimes result from interactions of polarons, and how local correlations in the polaronic and the Fermi-liquid regime are connected constitute essential questions of the high-$T_{c}$ puzzle.

A paradigmatic description of strongly-correlated quantum materials is the two-dimensional Fermi-Hubbard model. Despite recent progress in its numerical analysis \cite{LeBlanc2015, Chen2020}, a thorough understanding of this model is still lacking, which makes it a primary target for quantum simulation. The model consists of spin-$1/2$ fermions on a lattice with nearest-neighbour (NN) tunneling amplitude $t$ and on-site repulsion $U$, which leads to antiferromagnetic spin couplings $J$. Cold-atom based quantum simulators provide fully tunable implementations of such systems with single-site resolved readout and continuous doping control \cite{Gross2017}. Recent studies of systems in- and out-of-equilibrium started to characterize transport coefficients \cite{Nichols2018,Brown2019a} and two-point correlations \cite{Ji2020,Chiu2019,Hartke2020, Mazurenko2017} in doped Mott insulators. With the advent of full spin- and density resolution \cite{Boll2016, Koepsell2020}, spin-charge correlators enabled imaging of the dressing cloud of magnetic polarons \cite{Koepsell2019} and the exploration of spin-charge separation in one dimension \cite{Vijayan2019,Salomon2019, Hilker2017}. 

Here we study the hole-doping dependence of multi-point correlations between spin and charge (density) in two-dimensional Fermi-Hubbard systems and observe a simultaneous change across all presented observables around a specific doping $\delta_{\text{FL}}$, see Fig.~\ref{fig:fig1}A. Above $\delta_{\text{FL}}$ we identify the metal as a conventional Fermi liquid, while for lower dopings our experimental observables indicate a regime not captured within conventional perturbative and mean-field frameworks. We track the evolution of the polaronic dressing cloud of single holes and probe magnetic correlations surrounding hole pairs for interaction effects, offering new insight on this crossover beyond traditional solid state observables. Furthermore, we perform a detailed comparison to numerical calculations and benchmark three prominent approximate theories for the low doping physics, which become increasingly disinguishable with higher-order correlators.


In the experiment, we realized two-dimensional Fermi-Hubbard systems at strong interactions $U/t\ \sim 8$ using $^{6}$Li atoms in the lowest two hyperfine states in an optical lattice with spacing $a=1.15\,\mu$m as described in previous work \cite{Koepsell2020}. Full spin- and density readout is achieved by detecting each spin component separately in adjacent layers of a vertical superlattice \cite{Koepsell2020}, see Fig.~\ref{fig:fig1}B. The Gaussian envelope of our optical beams creates a harmonic trapping potential, which naturally leads to an increasing hole-doping from the center to the edge of our system. We use this spatial variation, together with our control of the total number of fermions in the system, to study the doping dependence of multi-point correlators \cite{SM}. To explore all relevant hole-doping regimes we use samples with up to $\sim100$ atoms and temperatures down to $k_{B}T=0.43(3)\,t$ (see \cite{SM}), where $k_{B}$ is the Boltzmann constant.

We study the connected part of bare $N$-point correlations, which contains the new information of order $N$ \cite{Schweigler2017} as illustrated in Fig.~\ref{fig:fig1}C. Bare correlations can arise from lower-order contributions (disconnected part), while the connected part measures genuine higher-order effects.

The numerics to which we compare are at finite temperature $k_{B}T=0.4\,t$ and can be divided into three categories, see Fig.~\ref{fig:fig1}D (see \cite{SM} for details on all calculations). Non-interacting (free) fermions and perturbation theory related methods are used to identify the FL regime at high doping. Two versions of Anderson's resonating-valence-bond (RVB) states \cite{Anderson1987}, namely uniform and $\pi$-flux, as well as a model for mutually independent magnetic polarons (string) are tested for their potential to capture low doping physics. Finally, exact diagonalization (ED) of finite size Fermi-Hubbard systems with $4\times4$ sites is included.
%

\begin{figure*}[]
\includegraphics{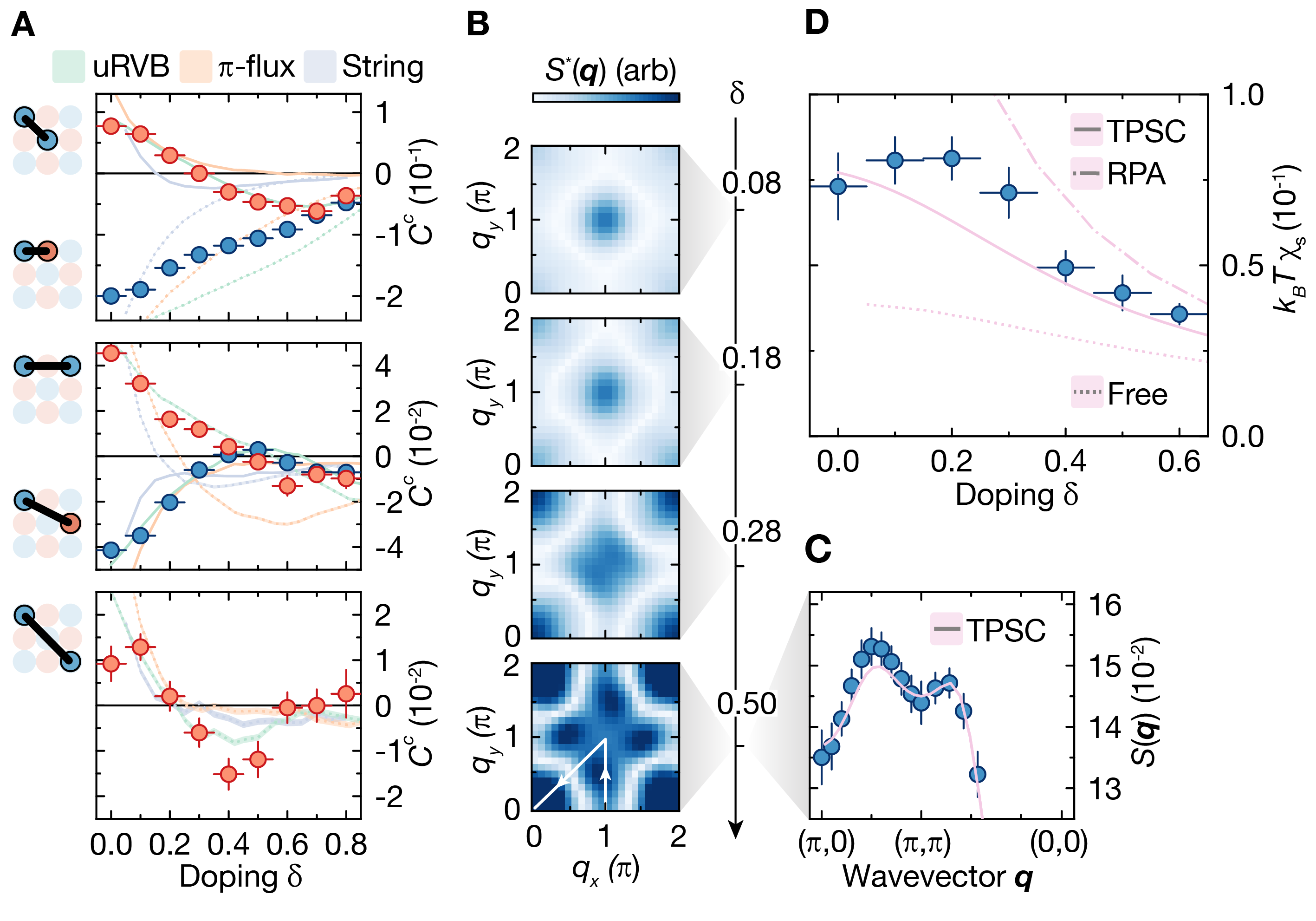}
\caption{\textbf{Magnetism from Mott insulator to Fermi liquid.} \textbf{A}, Connected two-point spin correlations as a function of doping for different spin distances (see insets). Error bars denote one standard error of the mean (s.e.m) and for doping the bin width for averaging. Solid (dotted) lines of numerical calculations indicated in the legend correspond to spin distances of red (blue) data points. Shaded bands indicate the statistical s.e.m. for all calculations where visible. \textbf{B}, Offset adjusted static spin-structure factor $S^{\star}(\boldsymbol{q})$ for increasing doping with arbitrary scales  and \textbf{C}, trace through unadjusted spin-structure factor $S(\boldsymbol{q})$ at $50\,\%$ doping. The full width of doping bins for \textbf{B}, \textbf{C} is 0.14. Solid pink represents a mean-field related TPSC calculation. \textbf{D}, Doping dependence of the uniform magnetic susceptibility, obtained via the fluctuation-dissipation relation. Solid, dashed and dotted pink curves correspond to TPSC, RPA and free fermion calculations (c.f. legend). This figure is based on $3\,224$ experimental realizations at $k_{B}T=0.43(3)\,t$ and $U/t=8.9(5)$.}
\label{fig:fig2}
\end{figure*}


First, we investigate how the antiferromagnetic alignment of two spins at positions $\boldsymbol{r}_1,\boldsymbol{r}_2$ evolves, by measuring connected two-point correlations (referred to as a bond)
\begin{equation}
C^{c}(\boldsymbol{d})=C^{c}(\boldsymbol{r}_1,\boldsymbol{r}_2)=\eta (\langle \hat{S}^{z}_{\boldsymbol{r}_{1}} \hat{S}^{z}_{\boldsymbol{r}_{2}} \rangle -\langle \hat{S}^{z}_{\boldsymbol{r}_{1}} \rangle \langle \hat{S}^{z}_{\boldsymbol{r}_{2}} \rangle),
\label{eq:eq1}
\end{equation}
where the normalization $\eta=1/(\sigma(\hat{S}^{z}_{\boldsymbol{r}_{1}})\sigma(\hat{S}^{z}_{\boldsymbol{r}_{2}}))$ yields a universal quantification of the correlation and $\sigma$ denotes the standard deviation. In the Heisenberg limit at half filling $\eta=4$. The bond length, which is the distance between two spins, is given by $\boldsymbol{d}=\boldsymbol{r}_2-\boldsymbol{r}_1$. As shown in Fig.~\ref{fig:fig2}A, doping quickly reduces the amplitude of antiferromagnetic correlations and leads to weakly oscillatory behavior as a function of doping. Between $\delta\sim$ $20-40\,\%$ spin correlations at different distances (such as $d=\sqrt{2},2,\sqrt{5},\sqrt{8}$) undergo a sign reversal. The uniform-RVB state features similar sign flips of correlations and compares well also for larger dopings. $\pi$-flux and the string model behave similarly and show agreement with our data for $\delta<20\,\%$, in line with \cite{Chiu2019}. Predictions for two-point correlations of different theoretical approaches are very similar at low doping, which calls for a comparison of higher-order spin-charge correlations.

Above hole concentrations around $50\,\%$, oscillating magnetism manifests itself as visible peaks in the static spin-structure factor $S(\boldsymbol{q})$ shifting from $(\pi,\pi)$ towards $(\pi,0)$. The effect is even more pronounced in an adjusted version $S^{\star}(\boldsymbol{q})$, which neglects the strong on-site term $d=0$ equivalent to a broad offset in Fourier space, see Fig.~\ref{fig:fig2}B,C and \cite{SM}. This shift of fluctuations towards momenta incommensurate with the lattice spacing is in excellent agreement with a perturbation theory inspired two-particle-self-consistent approach (TPSC) \cite{Vilk1994} and confirms Quantum-Monte-Carlo (QMC) calculations \cite{Moreo1990, Furukawa1992}. This indicates, the observed shift of spin fluctuations can be considered a Fermi-liquid phenomenon, where a stretch of the Fermi wavevector $\boldsymbol{q}_{\text{F}}$ with increasing doping causes such incommensurate fluctuations through interactions on a mean-field level. A possible connection to incommensurate spin-density wave phases (stripes) at weak doping and colder temperatures \cite{Cheong1991} needs further exploration. 

\begin{figure*}[]
\includegraphics[scale=1]{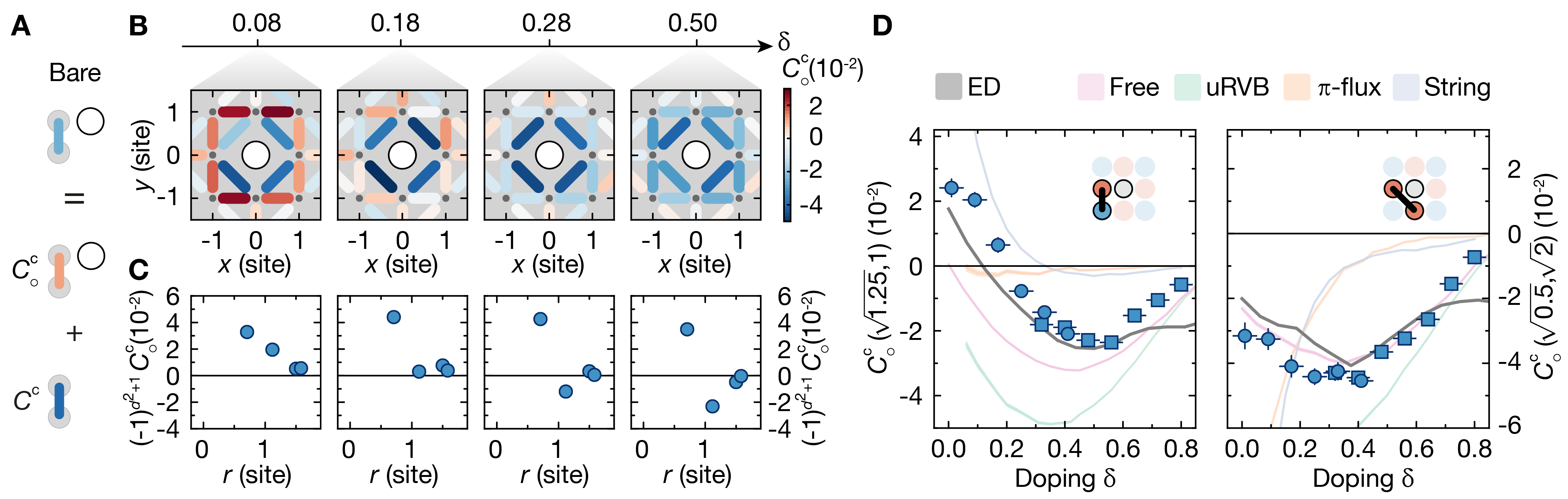}
\caption{\textbf{Breakdown of polaronic correlations.} \textbf{A}, Relation between bare and connected spin correlations in the vicinity of a hole. \textbf{B}, Connected correlation (represented as bonds) of spins on NN and diagonal lattice sites (grey dots) in the presence of a single hole (white central dot) for different dopings. \textbf{C}, Connected correlations as a function of bond distance $r$ from the hole, where we flip the sign of correlations with bond length $d=\sqrt{2}$. Thus a positive correlation indicates a connected signal opposing the two-point correlations at half filling. Error bars denote one s.e.m. and are smaller than the point size. The full width of doping bins for \textbf{B}, \textbf{C} is 0.1. \textbf{D}, Doping dependence of the NN and diagonal bonds closest to the hole (see insets). Square (circular) datapoints were extracted from a dataset with $52.0(1)$ ($91.3(1)$) average number of particles. Solid lines represent numerical calculations (see legend) and shaded bands indicate (where visible) their statistical s.e.m. This figure is based on $18\,107$ experimental realizations at $k_{B}T=0.52(5)\,t$ and $U/t=7.4(8)$.}
\label{fig:fig3}
\end{figure*}

Furthermore, we extract the doping dependence of the uniform ($\boldsymbol{q}=\textbf{0}$) spin-susceptibility $\chi_{\text{s}}$, see Fig.~\ref{fig:fig2}D, by applying the fluctuation-dissipation relation \cite{SM} in an approach similar to \cite{Hartke2020,Drewes2016}. We compare experimental data to three FL type calculations: free fermions without interaction, a random-phase-approximation (RPA) at lower effective $U/t=4$ to avoid divergences (see \cite{SM}) and TPSC. For $\delta>\delta_{\text{FL}}\sim30\,\%$ the susceptibility increases with decreasing doping, which is quantitatively best captured by TPSC calculations. However, below $\delta_{\text{FL}}$ the susceptibility $\chi_{\text{s}}$ stops increasing for weaker dopings. This behavior is reminiscent of the pseudogap phenomenon as well as anomalous with respect to our FL calculations, and supported by QMC results \cite{Moreo1993}. This indicates, that the metallic regime below $\delta_{\text{FL}}$ is of a different nature than the conventional Fermi liquid found at higher dopings (for convergence of structure factors in FL see \cite{SM}).

The weakly doped metallic regime hosts magnetic polarons, whose dressing cloud can be measured with a three-point correlator of two spins around a hole \cite{Koepsell2019, Blomquist2019}. For spin-balanced systems $\langle \hat{S}^{z}_{\boldsymbol{r}_i} \rangle=0$, the connected part simplifies to \cite{SM}
\begin{align}
C^{c}_\circ(\boldsymbol{r},\boldsymbol{d})=C^{c}_\circ(\boldsymbol{r}_3;\boldsymbol{r}_1,\boldsymbol{r}_2)= 
\\
\eta \langle\hat{S}^{z}_{\boldsymbol{r}_{1}} \hat{S}^{z}_{\boldsymbol{r}_{2}} \rangle_{\circ _{\boldsymbol{r}_3}}-C^{c}(\boldsymbol{r}_1,\boldsymbol{r}_2) \nonumber
\label{eq:eq2}
\end{align} 
and measures how the bond is perturbed away from the background two-point correlation by post-selecting on a hole at a third position $\boldsymbol{r}_3$, c.f. Fig.~\ref{fig:fig1}C. The distance of the bond center to the hole is given by $\boldsymbol{r}=(\boldsymbol{r}_1+\boldsymbol{r}_2)/2-\boldsymbol{r}_3$. 

For $\delta$ around $10\,\%$, a hole perturbs all bonds in its vicinity with a sign opposite to the antiferromagnetic background, such that NN spins ($d=1$) align more ferromagnetically (parallel) and diagonal spins ($d=\sqrt{2}$) more antiferromagnetically (antiparallel), see Fig.~\ref{fig:fig3}. Doublon-hole fluctuations cause a similar connected signal already at half-filling, but play a minor role at $10\, \%$ doping \cite{SM}. When measuring the strength of this effect versus bond distance from the hole, the radial dependence of the polaronic dressing is obtained (see Fig.~\ref{fig:fig3}B). 

In the Fermi-liquid regime at large doping, the Pauli exclusion principle prevents fermions with the same spins to occupy sites in a small volume \cite{Hartke2020}. This causes an enhanced antiferromagnetic alignment of all bonds (also $d=1$) in the presence of a hole and in fact is expected to cause small amplitude oscillations of that alignment with larger distance from the hole, akin to Friedel oscillations around a static hole. 

Therefore, a useful indicator for the transition between the two metals is the NN bond ($d=1$) closest to the hole, whose connected correlation continuously evolves from ferromagnetic to antiferromagnetic across the regimes, see Fig.~\ref{fig:fig3}C. An intial drop of the connected signal is expected from the higher concentration of polarons, as their dressing clouds start to overlap. Around $20\,\%$ doping, the closest NN bond becomes uncorrelated with the presence of the hole and builds up an antiferromagnetic alignment towards $\delta_{\text{FL}}$, consistent with ED. At a similar doping $\delta \sim \delta_{\text{FL}}$, the closest distance connected diagonal correlations are maximally antiferromagnetic.

String and RVB predictions for $C^{c}_{\circ}$ are very distinguishable at weak dopings. Only the polaron model (string) reproduces the experimental ferromagnetic alignment of the closest NN bond, while RVB states show strong discrepancies to experiment. Uniform RVB is a prime example of how a theoretical approach can show excellent agreement with experiment in two-point correlations at low doping, but reveal strong deviations at higher-order correlators. At large dopings, uniform RVB and free fermions start to capture the correlations driven by fermionic statistics. 

QMC studies of Fermi-Hubbard systems found the bandwidth of quasiparticle excitations evolves from polaronic (order $2J$) to Fermi liquid (order $8t$) at around $30\,\%$ doping \cite{Preuss1997}. Our measurements suggest polaronic dressing persists up to $\delta \sim20\,\%$ and smoothly dissolves into Fermi-liquid correlations around $\delta_{\text{FL}}\sim30\,\%$.

\begin{figure}[]
\includegraphics[scale=1]{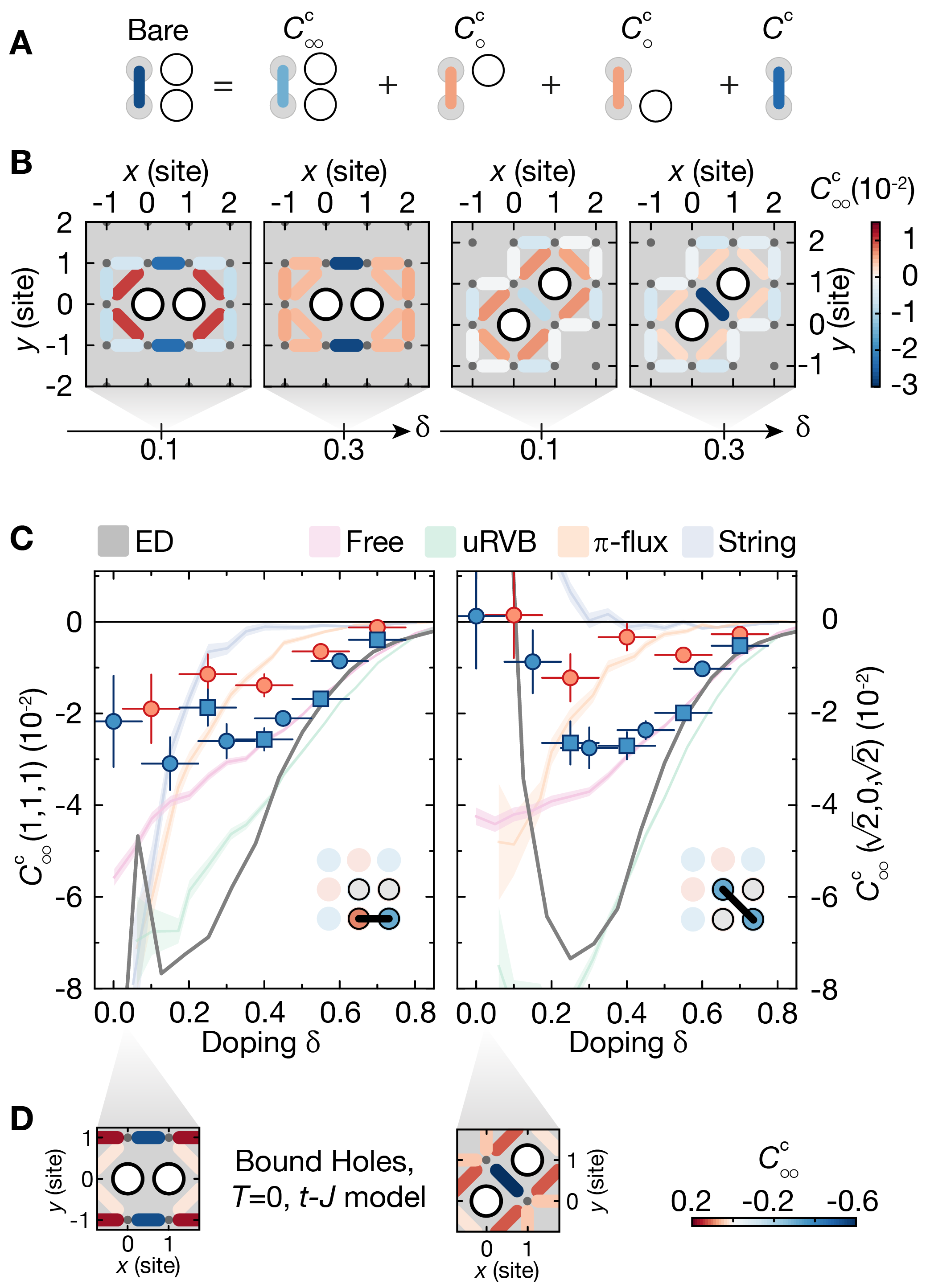}
\caption{\textbf{Influence of two holes on spin correlations.} \textbf{A}, Relation between bare and connected spin correlations in the vicinity of two holes. \textbf{B}, Connected correlations of NN and diagonal spins in the presence of a NN or diagonal pair of holes at two dopings with a full width of the doping bin of $0.2$. Same bond distances and symmetric spatial hole orientations are averaged together (see text). \textbf{C}, Connected correlation of the bond with closest distance to the NN or diagonal hole pair (see insets) as a function of doping. Blue (red) point correspond to experimental temperatures of $ 0.52(5)\,t$  ($ 0.77(7)\,t$). Blue square (circular) datapoints were extracted from a dataset with $52.0(1)$ ($91.3(1)$) average number of particles. Solid lines represent numerical calculations as indicated (see legend) and shaded bands their statistical standard error of the mean. \textbf{D}, DMRG calculations ($T=0$) for two holes in the $6$-leg ladder $t$-$J$ model, where binding occurs. This figure is based on $23\,695$ experimental realizations at $U/t=7.4(8)$.
}
\label{fig:fig4}
\end{figure}


When two polarons come close, their dressing clouds overlap, which can lead to the breakdown of polarons or induce effective interactions between them. This is often considered as a possible mechanism for pseudogap behavior \cite{Schrieffer1989, Keimer2015, Dagotto1994}. Hole-hole correlators do not show indications of hole binding at current temperatures of cold-atom quantum simulators \cite{Koepsell2019, Chiu2019, SM}, hence we search for interaction signatures in the magnetic environment of two holes. 

In the analysis, we post-select on two holes at positions $\boldsymbol{r}_3,\boldsymbol{r}_4$ and evaluate the connected (four-point) correlation between two spins in the presence of a hole pair, which in a spin-balanced system reduces to
\begin{align}
C^{c}_{\circ \circ}(\boldsymbol{l},\boldsymbol{r},\boldsymbol{d})=C^{c}_{\circ \circ}(\boldsymbol{r}_3,\boldsymbol{r}_4;\boldsymbol{r}_1,\boldsymbol{r}_2)=\eta \langle\hat{S}^{z}_{\boldsymbol{r}_1} \hat{S}^{z}_{\boldsymbol{r}_2} \rangle_{\circ_{\boldsymbol{r}_3}\circ_{\boldsymbol{r}_4}}\\-C^{c}(\boldsymbol{r}_1,\boldsymbol{r}_2)-
 \gamma (C^{c}_{\circ}(\boldsymbol{r}_3;\boldsymbol{r}_1,\boldsymbol{r}_2)+C^{c}_{\circ}(\boldsymbol{r}_4;\boldsymbol{r}_1,\boldsymbol{r}_2)) \nonumber,
 \label{eq:eq3}
\end{align}
see Fig.~\ref{fig:fig4}A (for the general expression see \cite{SM}). The mutual distance of the holes is defined as $\boldsymbol{l}=\boldsymbol{r}_4-\boldsymbol{r}_3$ and the bond distance $\boldsymbol{r}$ is measured w.r.t. the center of $\boldsymbol{l}$.
$C^{c}_{\circ \circ}$ detects correlations linked to the presence of the holes as a pair and measures how much these deviate from a simple addition of two independent single-hole signals $C^{c}_{\circ}$ with a weighting factor $\gamma=\langle \hat{h}_{\boldsymbol{r}_3} \rangle \langle \hat{h}_{\boldsymbol{r}_4} \rangle/\langle \hat{h}_{\boldsymbol{r}_3} \hat{h}_{\boldsymbol{r}_4} \rangle$ and hole density operator $\hat{h}_{\boldsymbol{r}_i}$. 

We study the case of NN ($l=1$) or diagonal ($l=\sqrt{2}$) hole pairs and bonds $d=1,\,\sqrt{2}$. To obtain a sufficient signal-to-noise ratio in the experiment we combine the two configurations for NN ($\boldsymbol{l}=(1,0),\,(0,1)$) and diagonal pairs ($\boldsymbol{l}=(1,1),\,(1,-1)$) by averaging all bonds with identical bond distance $r$ from the pair. To visualize correlations we choose a representation in terms of $\boldsymbol{l}=(1,0)$ and $\boldsymbol{l}=(1,1)$, see Fig.~\ref{fig:fig4}B. We find connected antiferromagnetic alignment of bonds at closest distance to the pair, which connects both metallic regimes. As shown in Fig.~\ref{fig:fig4}B,C, for NN holes the closest bond has a negative correlation at half filling (inherited from doublon-hole pairs \cite{SM}), which stays antiferromagnetic for higher doping and quantitatively agrees with Fermi-liquid correlations for $\delta>\delta_{\text{FL}}$. This bond is furthermore robust against an increase in temperature to $k_{B}T=0.77(7)\,t$. For diagonal holes, the diagonal spin bond between them has the shortest distance to the pair, see Fig.~\ref{fig:fig4}B,C. This bond is uncorrelated at half filling (doublon-hole pairs contribute a ferromagnetic signal, see ED at $\delta=0\,\%$ or \cite{SM}), then rapidly turns antiferromagnetic with doping, peaks at $\delta_{\text{FL}}\sim30\,\%$ and is eventually described quantitatively by Fermi-liquid correlations for $\delta>\delta_{\text{FL}}$. For higher temperatures, the correlation of this bond is significantly reduced. Approximate theories for low doping partly predict such antiferromagnetic correlations of closest distance bonds, but show limited overall agreement to experimental data.

To gain an intuition of how such correlations would connect to lower temperature physics, we consider two holes ($\delta\sim2\,\%$) in the $t$-$J$ model, for which binding of polarons (holes) occurs at relatively high temperatures \cite{Blomquist2020}. We performed density-matrix-renormalization-group (DMRG) calculations of this scenario at $T=0$ for a $6$-leg ladder \cite{SM} and show the connected spin environment in Fig.~\ref{fig:fig4}D for $l=1$ and  $l=\sqrt{2}$. A striking effect of hole pairing is the emergence of a strong antiferromagnetic spin bond at closest distance to the pair \cite{Blomquist2020, White1997a}. Our experimental correlations feature similar signatures, but no further indication of hole binding (see hole-hole correlations in \cite{SM}). This leads us to the conclusion, that qualitative features of the zero temperature physics of two holes are already encoded in the finite temperature limit and a strong interplay of spin and charge correlations already precedes hole pairing or formation of other competing orders at colder temperatures.


We harnessed the unique capability of our quantum simulator to study the continuous doping dependence of observables unavailable in traditional solid-state experiments and discovered a metal of magnetic polarons at weak doping and a Fermi liquid beyond $\delta_{\text{FL}}\sim30\,\%$. Their transition is signaled across all studied system properties (for a summarizing table see \cite{SM}) and the intricate spin-charge correlations reported serve as a novel basis to develop a microscopic understanding of pseudogap or collective phenomena at colder temperatures. How the observed doping for this crossover in our experiment can be related to solid-state measurements is unclear, since details like band structure and the difference in accessed observables plays an important role. In a benchmark of three approximate low doping theories, we find limited overall agreement with our system, calling for more efficient descriptions. Spin-charge correlators could also be studied in systems out-of-equilibrium \cite{Vijayan2019, Ji2020} and only modest improvements in colder temperatures with available cooling proposals \cite{Kantian2016} might enable experimental observation of pairing \cite{Blomquist2020} and pseudogap behavior \cite{Khatami2011}. Future studies could focus on fifth-order \cite{Bohrdt2020} correlators to further inspire our understanding of exotic many-body phenomena and test different theories \cite{Punk2015,Zhang2020}.

\bigskip
\begin{acknowledgments}
\textbf{Acknowledgments:} The authors would like to thank T.A. Hilker and T. Chalopin for insightful discussions and careful reading of the manuscript. This work was supported by the Max Planck Society (MPG), the European Union (FET-Flag 817482, PASQUANS), the Max Planck Harvard Research Center for Quantum Optics (MPHQ) and under Germany's Excellence Strategy  -- EXC-2111 -- 390814868. This research used resources of the National Energy Research Scientific Computing Center (NERSC), a U.S. Department of Energy Office of Science User Facility operated under Contract No. DE-AC02-05CH11231. J.K. gratefully acknowledges funding from Hector Fellow Academy. E.D. and Y.W. acknowledge support from Harvard-MIT CUA, ARO grant number W911NF-20-1-0163, and the National Science Foundation through grant No. OAC-1934714.

\textbf{Author contributions}: All authors contributed significantly to the work presented in this manuscript

\textbf{Competing interests}: The authors declare no competing interests.
\end{acknowledgments}

\setcounter{figure}{0}
\setcounter{equation}{0}
\renewcommand\thefigure{S\arabic{figure}}  
\renewcommand\thetable{S\arabic{table}} 
\renewcommand{\theequation}{S\arabic{equation}}%

\section{Supplementary Material}
\subsection{Data acquisition and characterization}
We prepared balanced cold atomic samples in the lowest two hyperfine states of $^{6}$Li, closely following our previous work \cite{Koepsell2020}. During evaporation, the gas was harmonically trapped in the $xy$-plane and vertically confined in a single layer of an optical superlattice with lattice spacings $a_s=3\,\mu$m ($a_{l}=6\,\mu$m) and depths $50\,E_{R}^{s}$ ($100\,E_{R}^{l}$), where $E_{R}^{i}$ denotes the recoil energy of the respective lattice. The superlattice was set to a maximally tilted double-well configuration and atoms were initialized in the lower well before evaporation. The final particle number was controlled by the evaporation parameters. After evaporation, a two-dimensional $xy$-lattice with spacings $a_{x}=a_{y}=a=1.15\,\mu$m was ramped to around $6.5\,E_{R}^{xy}$ within $100\,$ ms and the scattering length was tuned to $810\,a_{B}$, where $a_{B}$ is the Bohr radius, using the broad Feshbach resonance of $^{6}$Li. For detection, spin-resolution was achieved by the method presented in \cite{Koepsell2020} and single-site resolved fluorescence images were taken in a dedicated pinning lattice \cite{Omran2015}. 

\begin{figure}[b]
\includegraphics{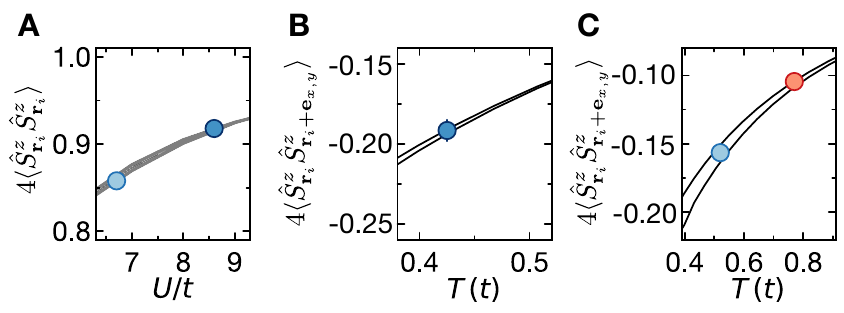}
\caption{\textbf{Extraction of temperature and consistency check of interaction strength of datasets.} \textbf{A}, NLCE calculations (solid black) of on-site fluctuations from reference \cite{Khatami2011} versus interaction strength at half filling for temperatures of $k_{B}T/t=0.76$ (lower line) and $k_{B}T/t=0.40$ (upper line). On-site fluctuations are almost temperature independent and can be used to extract the interaction strength and confirm our system calibration (see text)	. Datasets (D1,D2) are (dark,light) blue points and are consistent with $U/t$ of $(8.6,\,6.7)$. \textbf{B}, \textbf{C}, NLCE calculations (solid black) of nearest-neighbour spin correlations at half filling versus temperature. The (upper, lower) curves are at $U/t=(7,9)$ in \textbf{B} and $U/t=(6,8)$ in \textbf{C}. Points in (dark blue, light blue, red) correspond to datasets (D1,D2,D4) and error bars denote one s.e.m. For experimental data a density filter of $[0.96,1.03]$ was used for dataset D1 and $[0.92,0.97]$ for datasets D2, D3, D4.}
\label{fig:figsi1}
\end{figure}

Four datasets [D1,D2,D3,D4] with a total of $[3224,\, 8667,\, 9440,\, 5588]$ realizations were taken. The final $x$- and $y$-lattice depths for dataset D1 were $(6.9\,E_{R}^{x},\,6.9\,E_{R}^{y})$. For datasets [D2,D3,D4] the $xy$-lattice spacings were slightly different $a_{x}/a_{y}=1.02$ and therefore final lattice depths were chosen to be $(6.5\,E_{R}^{x},\,6.7\,E_{R}^{y})$ to yield symmetric tunneling elements $t_{x}=t_{y}$. The short spaced vertical lattice was $50\,E_{R}^{s}$ for [D1] and $44\,E_{R}^{s}$ for [D2,D3,D4]. We performed a Wannier function calculation to estimate the absolute tunneling amplitude for settings of datasets [D1] and [D2,D3,D4] to be $t/h=240(10)\,$Hz and $t/h=260(10)\,$Hz. The mean particle numbers of the four datasets are $[89.8(1),\,91.3(1),\,52.0(1),\,90.8(1)]$. For D4, atoms were held in the harmonic trap for $1.75\,$s before loading the $xy$-lattice to produce systems at a higher temperature. 

The single-particle detection fidelity $p$ for datasets is slightly different and estimated to be $p=97\,\%$ for D1 and $p=95\,\%$ for [D2,D3,D4] by comparing occupations in subsequent images of the same realization. We do not renormalize observables by this fidelity, except for the temperature and interaction extraction (see below). A possible renormalization of observables by this fidelity would not lead to any significant change of results presented in this work. Error bars for all correlator-based observables were found by performing a bootstrap and computing the standard deviation of the mean across the resampled datasets.

The figures $2,3,4$ of the main manuscript are based on datasets [D1], [D2,D3], [D2,D3,D4].

\subsection{Interaction strength and temperature}

We estimate the interaction strength by a Wannier function calculation, given our calibrated system parameters. This yields $U/t=9.3$ for D1 and $U/t=8.2$ for [D2,D3,D4]. A comparison of the on-site fluctuations $4\langle \hat{S}^{z}_{\boldsymbol{r}_{i}} \hat{S}^{z}_{\boldsymbol{r}_{i}} \rangle$ at half filling to numerical linked cluster expansion (NLCE) calculations of reference \cite{Khatami2011} is consistent with $U/t=8.6$ for D1 and $U/t=6.7$ for [D2,D3,D4] when corrected for our detection fidelity, see Fig.~\ref{fig:figsi1}A. We therefore combine our calibration and information from NLCE to assess the interaction strength to be $8.9(4)$ for D1 and $7.4(8)$ for [D2,D3,D4].

We extract the temperature $T$ of datasets [D1,D2,D4], by comparing the nearest-neighbour spin correlation $4\langle \hat{S}^{z}_{\boldsymbol{r}_{i}} \hat{S}^{z}_{\boldsymbol{r}_{i}+\textbf{e}_{x,y}} \rangle$ at half filling with (NLCE) calculations, where we average correlations with $\textbf{e}_{x}=(1,0)$ and $\textbf{e}_{y}=(0,1)$. As shown in Fig.~\ref{fig:figsi1}B,C and taking into account our detection fidelity, we find the datasets [D1,D2,D4] are consistent with temperatures $k_{B}T/t$ of $[0.43(3),\,0.52(5),0.77(7)]$. Temperate uncertainties are estimated by the best- and worst-case scenarios, given the statistical errorbar of the spin correlation and the uncertainty in $U/t$. The temperature of dataset D3 is equal to D2,  as their spin correlation strength coincides at the same doping and their experimental sequence only differs by total particle number. 

\subsection{Doping analysis}
\begin{figure}
\includegraphics[width=0.48 \textwidth]{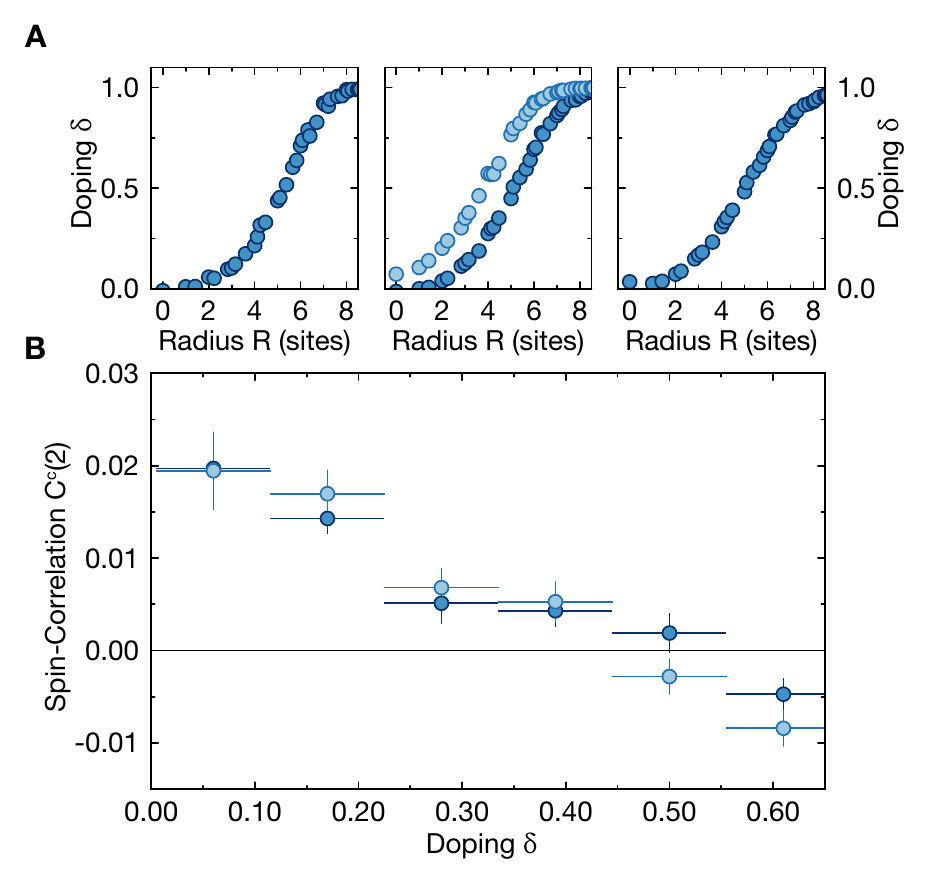}
\caption{\textbf{Spatial distribution of doping.} \textbf{A}, Radially averaged hole-doping concentration for all four datasets from center to edge of the system. Error bars denote one s.e.m. and are smaller than point size. \textbf{B}, Due to spatial doping gradients, we assign each local correlation the mean doping of all its points. The doping dependence of the correlation of two spins at intermediate distance $d=2$ is shown, calculated with this method from two datasets with different spatial doping distribution (different chemical potential). Both higher (dark blue) and lower (light blue) chemical potential agree qualitatively and quantitatively for this correlation.}
\label{fig:figsi2}
\end{figure}

The harmonic confinement of our atoms leads to increasing hole doping from the center to the edge of our systems. The radial dependence of the hole-doping concentration is shown for all four datasets in Fig. \ref{fig:figsi2}A. $N$-point correlators (in this work $N\in [2,3,4]$) locally extend over $N$ lattice sites, of which not all share the same doping concentration, due to the spatial doping gradient in the system. When we compute the local value of a correlator, we label the calculated correlation value by the mean density of all its contributing $N$ sites $n=\sum n_i/N$ and therefore a doping $\delta = 1-n$. We average all local correlations with an assigned doping within a bin of width $\Delta \delta$ and centered around $\delta_c$, such that $\delta \in [\delta_c-\Delta \delta/2,\delta_c+\Delta \delta/2]$ to obtain the doping dependence of various correlations. In the analysis, we display the averaged value at a doping $\delta_c$ with an error bar of width $\Delta \delta$. The validity of this approach in our experimental system is supported by the agreement of the doping dependence of all our observables when compared between two datasets with different total particle numbers. In addition to the agreement already displayed in the main text figure Fig.~3 and Fig.~4, we show in Fig. \ref{fig:figsi2}B another example of this agreement. The correlation of spins at distance $d=2$ apart from each other displays the same quantitative and qualitative doping dependence  for the slightly and more heavily doped dataset. With more homogeneous systems obtained through potential shaping in the future, correlators occupying a larger spatial area or a more precise doping resolution will become accessible.

\begin{figure}
\includegraphics[width=0.48 \textwidth]{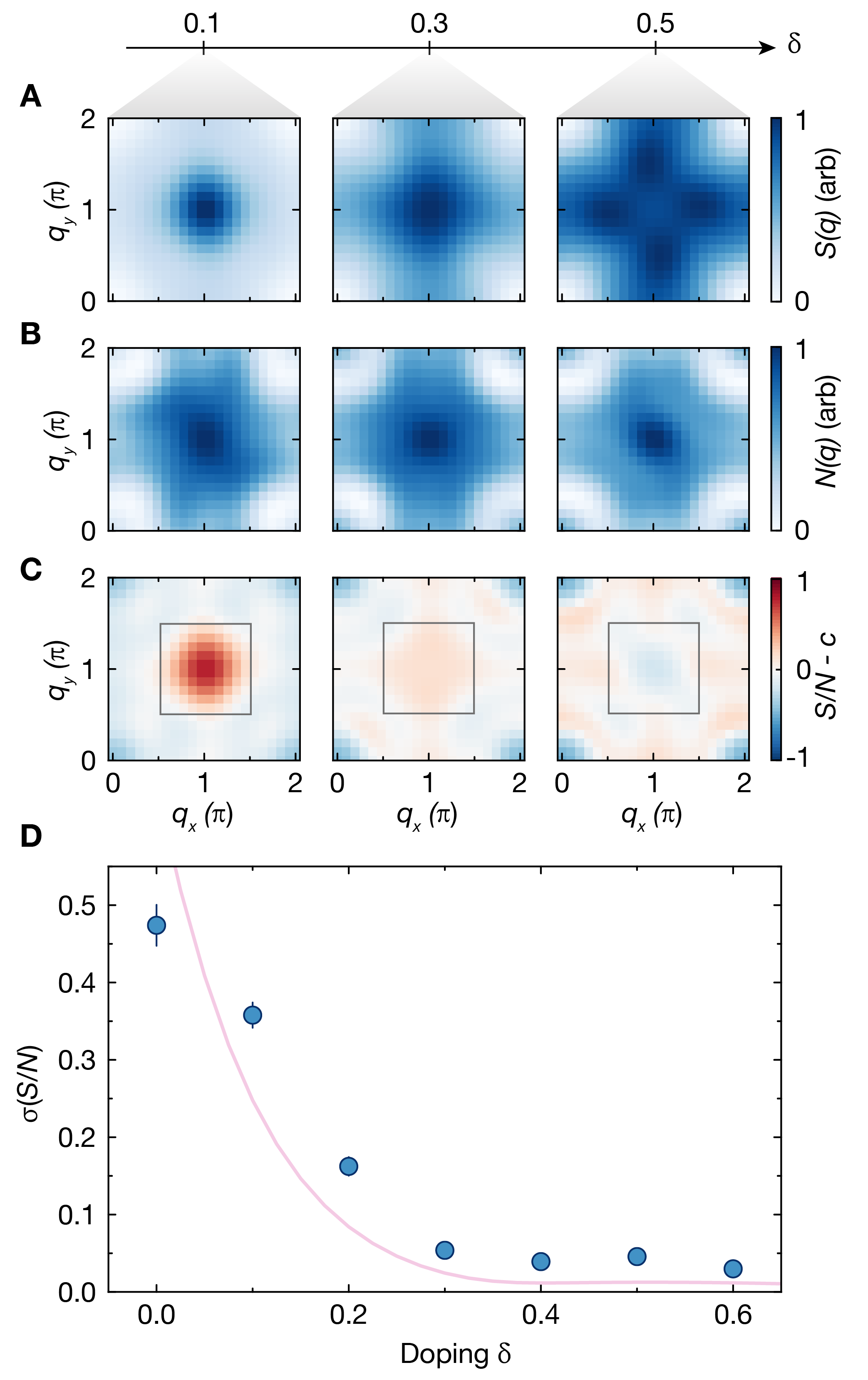}
\caption{\textbf{Spin and density structure factors.} \textbf{A}, Spin structure factor $S(\boldsymbol{q})$, \textbf{B}, density structure factor $N(\boldsymbol{q})$ and \textbf{C}, their ratio $S/N-c$, where $c$ is the mean ratio across all $\boldsymbol{q}$, for different doping levels (see insets). The total width of the doping bins is $0.15$. \textbf{D}, Standard deviation of the ratio $S/N$ across the momentum area (gray square) indicated as inset in \textbf{C} as a function of doping. Solid pink denotes our TPSC calculation at the system parameters. }
\label{fig:convergence}
\end{figure}

\subsection{Spin structure factor and susceptibility}
For our temperatures, the spin correlation length is short enough to approximate the thermodynamic limit (infinitely large system) of the structure factor with short distance correlations. We compute the static spin structure factor $S(\boldsymbol{q})=\sum_{d= 0}^{d=d_{c}} \langle \hat{S}^{z}_{\boldsymbol{r}_i}\hat{S}^{z}_{\boldsymbol{r}_i+\boldsymbol{d}}\rangle e^{i\boldsymbol{q}\boldsymbol{d}}$, based on an implicit average of spin correlations with $\boldsymbol{r}_i$ at a selected doping concentration and with a cutoff at maximal distance $d_c=\sqrt{10}$. If all neglected distances $d>\sqrt{10}$ have vanishing correlation values, this structure factor estimates the thermodynamic limit. Since the correlation $\langle \hat{S}^{z}_{\boldsymbol{r}_i} \hat{S}^{z}_{\boldsymbol{r}_i+\boldsymbol{d}} \rangle$ falls off with increasing distance $d$ at our temperatures, the contribution of distances $d>\sqrt{10}$ to the structure factor is indeed negligible compared to the much stronger shorter distances (at half filling $C^{c}(d=\sqrt{13})=-0.005(3)$). We keep a high number of points in momentum space, by padding distances up to $d=14$ with a correlation value of zero, which does not add nor affect any information encoded in our Fourier observables. To remove a constant and broad offset in momentum space, we exclude the strong positive on-site term $d=0$ from the Fourier transform and calculate $S^{\star}(\boldsymbol{q}) = \sum_{d=1}^{d=d_c} C^{c}(\boldsymbol{d})e^{i \boldsymbol{q}\boldsymbol{d}}$, which furthermore differs from $S(\boldsymbol{q})$ by the doping-dependent renormalization $\eta$ as defined in the main text. In main text Fig. 2B, $S^{\star}$ yields a cleaner signal of the incommensurate fluctuations, which is also confirmed in a cut through $S$ in Fig. 2C. We used $S(\textbf{0})$ to measure the doping dependence of the uniform magnetic susceptibility via the fluctuation-dissipation relation $\chi_{\text{s}}(\boldsymbol{q}=\textbf{0})k_{B}T = \,S(\textbf{0})$ \cite{ColemanBook}. This relation holds in this form only for $\boldsymbol{q}=0$ and was used with density correlations in previous work \cite{Hartke2020,Zhou2011,Drewes2016}. Since the entire system is in equilibrium, all different dopings are at the same temperature $T$. 

\subsection{Convergence of structure factors in the Fermi-liquid regime}
In weakly interacting Fermi-liquids, the static charge structure factor $N(\boldsymbol{q})=\sum_{d=0}^{d=d_c} \langle \hat{n}_{\boldsymbol{r}_i}\hat{n}_{\boldsymbol{r}_i+\boldsymbol{d}}\rangle e^{i\boldsymbol{q}\boldsymbol{d}}$ and $S(\boldsymbol{q})$ should eventually become similar. $N(\boldsymbol{q})$ was obtained by a Fourier transform of density-density correlations, similar to the analysis of $S(\boldsymbol{q})$. We quantify the similarity of spin and density structure factors in the momentum area around $\boldsymbol{q}=(\pi,\pi)$ by $\sigma(S/N)=\sqrt{\langle (S(\boldsymbol{q})/N(\boldsymbol{q}))^{2}\rangle_{\boldsymbol{q}}-( \langle S(\boldsymbol{q})/N(\boldsymbol{q})\rangle_{\boldsymbol{q}})^{2}}$, where $\langle...\rangle_{\boldsymbol{q}}$ denotes an average over all $\boldsymbol{q}$ within an area $[(\pi/2,\pi/2)+(q_x,q_y)$, $\pi>q_x,q_y>0]$. When $\sigma(S/N)$ goes to zero, both structure factors are related by a scaling factor independent of $\boldsymbol{q}$ in the chosen momentum space area. An example of $S(\boldsymbol{q})$, $N(\boldsymbol{q})$ and their ratio is shown for three doping levels in Fig. \ref{fig:convergence}. As shown in Fig. \ref{fig:convergence}D, spin and density structure factors have reached good convergence towards each other at $\delta_{\text{FL}}\sim30\,\%$, in agreement with our TPSC calculation.

%
%
%
\begin{figure}
\includegraphics{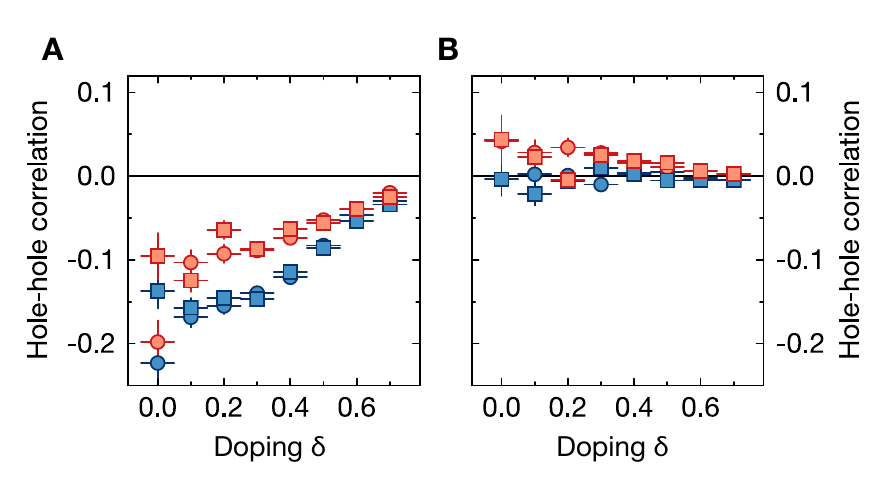}
\caption{\textbf{Hole-hole Correlations.} Doping dependence of $g^{(2)}_{hh}$-correlations (see text) between  \textbf{A}, NN and \textbf{B}, diagonal holes at $U/t=7.4(8)$. Blue (red) denote datasets at temperature $0.52(5)\,t$ and $0.77(7)\,t$. Circle, square data points represent \textbf{A}, $\boldsymbol{d}=(0,1),(1,0)$ and \textbf{B}, $\boldsymbol{d}=(1,1),(1,-1)$ directions.}
\label{fig:hhg2}
\end{figure}
\subsection{Hole-hole correlations}
Interactions between doped holes mediated by the spin background could manifest themselves as bunching of holes in real space, indicated by a positive correlation between two holes. At current accessible temperatures, we do not detect such an effect. In Fig. \ref{fig:hhg2}, we show the doping dependence of $g^{(2)}_{hh}=\langle \hat{h}_{\boldsymbol{r}_i}\hat{h}_{\boldsymbol{r}_j}\rangle/\langle\hat{h}_{\boldsymbol{r}_i} \rangle \langle \hat{h}_{\boldsymbol{r}_j} \rangle -1$ for NN and diagonal holes for two temperatures. Anti-correlation at the short distances considered here becomes stronger for colder temperatures.

\begin{figure}
\includegraphics{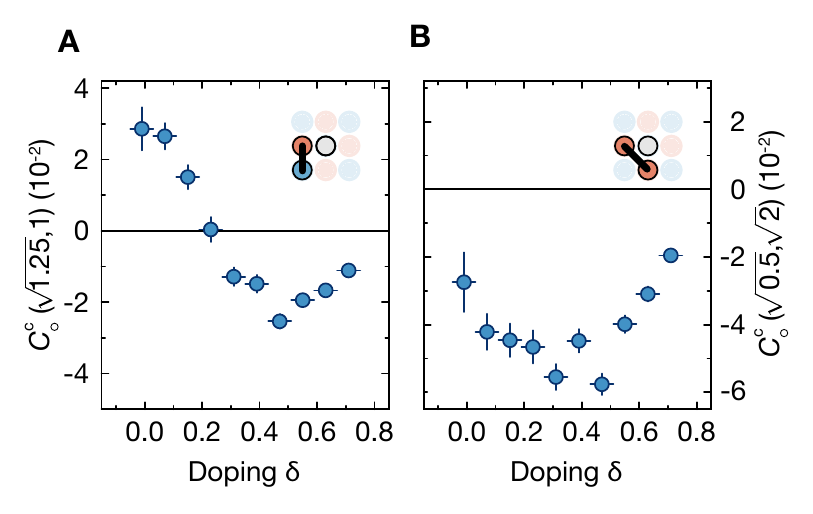}
\caption{\textbf{Connected hole-spin-spin correlations.} Connected correlation of \textbf{A}, NN  and \textbf{B} diagonal bonds at closest distance to the hole (see insets) for the dataset at $U/t=8.9(4)$.}
\label{fig:threepoint}
\end{figure}

\subsection{Connected correlator expressions}
The general ($\langle \hat{S}^{z}_{\boldsymbol{r}} \rangle \neq 0$) full expressions for the connected three- and four-point correlators presented in the main text are
\begin{align}
C^{c}_{3}(\boldsymbol{r}_1,\boldsymbol{r}_2,\boldsymbol{r}_3) = \langle \hat{h}_3 \hat{S}^{z}_2 \hat{S}^{z}_1 \rangle 
\\
- \langle \hat{h}_3 \rangle \langle \hat{S}^{z}_2 \hat{S}^{z}_1 \rangle
- \langle \hat{S}^{z}_2 \rangle \langle \hat{h}_3 \hat{S}^{z}_1 \rangle \nonumber
\\
- \langle \hat{S}^{z}_1 \rangle \langle \hat{h}_3 \hat{S}^{z}_2 \rangle + 2 \langle \hat{h}_3 \rangle \langle \hat{S}^{z}_2 \rangle \langle \hat{S}^{z}_1 \rangle \nonumber,
\end{align}

\begin{align}
C^{c}_{4}(\boldsymbol{r}_{1},\boldsymbol{r}_{2},\boldsymbol{r}_{3},\boldsymbol{r}_{4}) = \langle \hat{h}_4 \hat{h}_3 \hat{S}^{z}_2 \hat{S}^{z}_1 \rangle
\\
 - \langle \hat{h}_4 \rangle \langle \hat{h}_3 \hat{S}^{z}_2 \hat{S}^{z}_1 \rangle - 
\langle \hat{h}_3 \rangle \langle \hat{h}_4 \hat{S}^{z}_2 \hat{S}^{z}_1 \rangle  \nonumber
\\
- \langle \hat{h}_4 \hat{h}_3 \rangle \langle \hat{S}^{z}_2 \hat{S}^{z}_1 \rangle + 2 \langle \hat{S}^{z}_2 \hat{S}^{z}_1 \rangle \langle \hat{h}_4 \rangle \langle \hat{h}_3 \rangle \nonumber
\\
- \langle \hat{S}^{z}_2 \rangle \langle \hat{h}_4 \hat{h}_3 \hat{S}^{z}_1 \rangle - \langle \hat{S}^{z}_1 \rangle \langle \hat{h}_4 \hat{h}_3 \hat{S}^{z}_2 \rangle \nonumber
\\
-\langle \hat{h}_4 \hat{S}^{z}_2 \rangle \langle \hat{h}_3 \hat{S}^{z}_1 \rangle -\langle \hat{h}_4 \hat{S}^{z}_1 \rangle \langle \hat{h}_3 \hat{S}^{z}_2 \rangle \nonumber
\\
+2\langle \hat{h}_4 \hat{h}_3 \rangle \langle \hat{S}^{z}_2 \rangle \langle \hat{S}^{z}_1 \rangle + 2 \langle \hat{h}_4 \hat{S}^{z}_2 \rangle \langle \hat{h}_3 \rangle \langle \hat{S}^{z}_1 \rangle \nonumber
\\
 +2 \langle \hat{h}_4 \hat{S}^{z}_1 \rangle \langle \hat{h}_3 \rangle \langle \hat{S}^{z}_2 \rangle +2 \langle \hat{h}_3 \hat{S}^{z}_2 \rangle \langle \hat{h}_4 \rangle \langle \hat{S}^{z}_1 \rangle  \nonumber
 \\
+ 2 \langle \hat{h}_3 \hat{S}^{z}_1 \rangle \langle \hat{h}_4 \rangle \langle \hat{S}^{z}_2 \rangle - 6 \langle \hat{h}_4 \rangle \langle \hat{h}_3 \rangle \langle \hat{S}^{z}_2 \rangle \langle \hat{S}^{z}_1 \rangle \nonumber,
\end{align}
and their post-selected normalized forms
\begin{align}
C^{c}_{\circ}=C^{c}_{3}/(\langle \hat{h}_3 \rangle \sigma (\hat{S}^{z}_2) \sigma(\hat{S}^{z}_1)),
\end{align}

\begin{align}
C^{c}_{\circ \circ} = C^{c}_{4}/(\langle \hat{h}_4 \hat{h}_3 \rangle \sigma (\hat{S}^{z}_2) \sigma(\hat{S}^{z}_1)),
\end{align}
where we used an abbreviated subscript notation $\hat{O}_i$ for the operator $\hat{O}$ at position $\boldsymbol{r}_{i}$. For our analysis we always evaluated the full expression to avoid errors through possible small finite residual magnetizations.

\begin{figure}
\includegraphics[width=0.48 \textwidth]{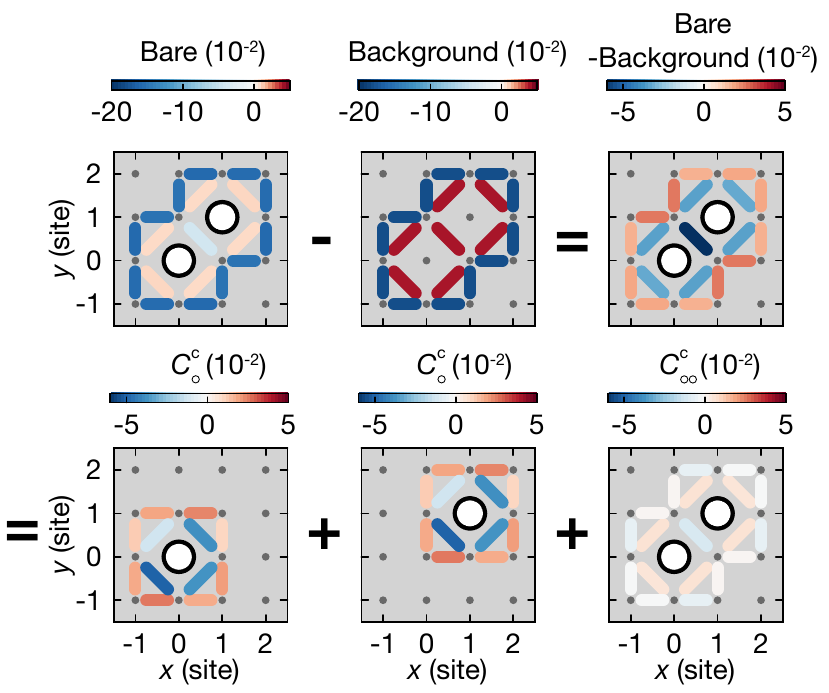}
\caption{\textbf{Decomposition of spin correlations surrounding a diagonal hole pair.} Bare $d=1,\sqrt{2}$ spin correlations in the presence of two holes and their difference from the bare strength of two-point spin correlations (background) can be decomposed into two independent connected three-point contributions ($C^{c}_{\circ}$) from each hole and higher-order effects measured by the connected four-point correlation for two holes ($C^{c}_{\circ\circ}$). The correlations shown are for doping of $\delta\in[0.05,0.15]$. The weighting factor $\gamma$ is experimentally close to $1$ and therefore neglected in this illustration.}
\label{fig:illustration}
\end{figure}

\begin{figure}
\includegraphics[width=0.48 \textwidth]{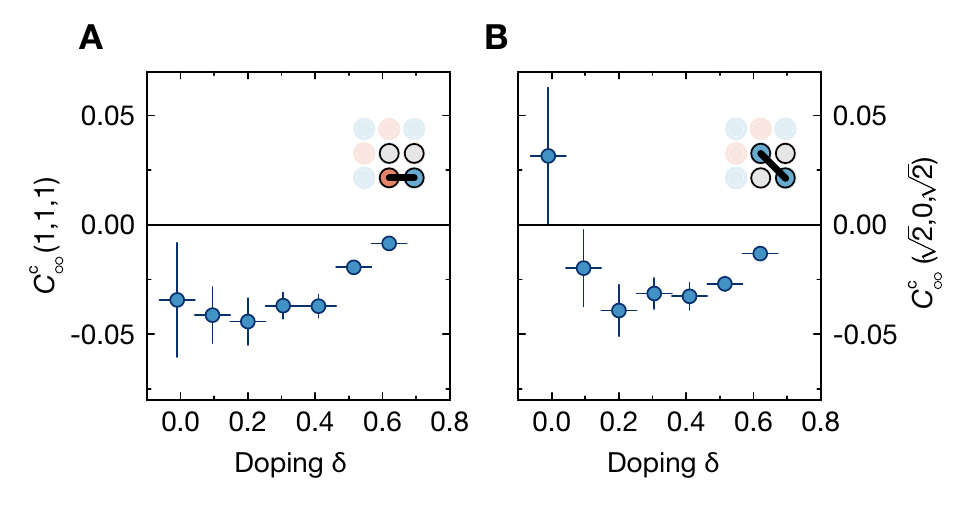}
\caption{\textbf{Connected hole-hole-spin-spin correlations.} Connected correlations of bonds with closest distance to \textbf{A}, NN and \textbf{B} diagonal pairs of holes (see insets) for the dataset at $U/t=8.9(4)$. Error bars for doping denote width of doping bin and one s.e.m. for correlation values.}
\label{fig:fourpoint}
\end{figure}

\subsection{Extended hole-spin-spin correlations}
Connected three-point correlations for the dataset at $U/t=8.9(4)$ and $k_{B}T=0.43(3)\,t$ are shown in Fig. \ref{fig:threepoint}. All findings from the main text can be verified also on this dataset.

\subsection{Extended hole-hole-spin-spin correlations}
An intuitive picture for the connected part can be gained when considering all contributions to the bare correlation. An illustration with experimental data is shown in Fig. \ref{fig:illustration}. In Fig. \ref{fig:fourpoint} we show the connected four-point correlations of the dataset at $U/t=8.9(4)$, which agrees with all observations from the main text.

\begin{figure}
\includegraphics[width=0.48 \textwidth]{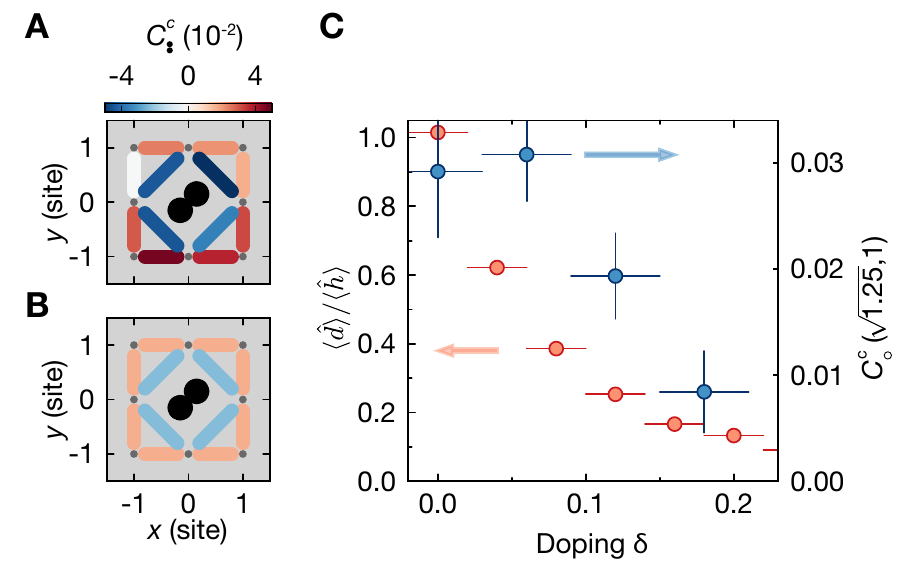}
\caption{\textbf{Influence of doublon-hole fluctuations on the spin environment.} \textbf{A}, Experimental connected spin correlations surrounding doublons at density $n \in [0.9,1]$, where all doublons originate from doublon-hole fluctuations. The effect on spin correlations is a reduction from the antiferromagnetic background, which is due to the hole on neighbouring sites. \textbf{B}, Exact diagonalization calculations of connected spin-spin-doublon correlations at half filling for $4\times4$ Fermi-Hubbard systems at $k_B T=0.4\,t$. \textbf{C}, Ratio between doublon-hole fluctuations and total number of holes (red) $\langle\hat{d}\rangle/\langle\hat{h}\rangle$, where $\hat{d}$ is the doublon-density operator, as a function of doping. Connected NN ($d=1$) spin correlations at closest distance from holes ($r=\sqrt{1.25}$) are overlayed (blue points) and show, that positive connected correlations persist also at dopings, where doublon-hole fluctuations are negligible.}
\label{fig:figsi4}
\end{figure}

\subsection{Influence of doublon-hole fluctuations}
In Fermi-Hubbard systems, short-range doublon-hole fluctuations exist at finite $U/t$, whereby a particle hops on top of a neighbouring one for a short time period despite the repulsive interaction $U$. This process is the strongest at half filling and is observed as doublons, which have a hole located mostly as a direct nearest neighbour. These holes are not distinguished from doped holes in our correlators. While the weight of their contribution is negligible compared to true holes in doped systems above $\delta\sim10\,\%$, their contribution becomes relevant for very low dopings close to half filling. The nearest-neighbour doublon of a hole belonging to such a fluctuation carries zero spin and therefore weakens the average antiferromagnetism around the hole. This is a different mechanism than the weakening of antiferromagnetism by a magnetic polaron, which is caused by a spinon bound to the hole in its immediate vicinity (in the string picture). In a similar manner, two holes, where each is part of a separate doublon-hole fluctuation, can have a nonzero connected four-point correlation with the spin environment.

The effect of doublon-hole fluctuations can be studied by investigating the connected three- and four-point correlations $C^{c}_{ \overset{\scalebox{0.45}{\newmoon}}{\scalebox{0.45}{\newmoon}}  } $ with doublons instead of holes at hole dopings close to half filling. In Fig. \ref{fig:figsi4} we show doublons belonging to doublon-hole pairs have a qualitatively similar connected correlation as found for holes at finite doping, which can be understood from the presence of a neighbouring hole for each doublon as explained above. When comparing the doublon-hole fluctuation concentration with connected correlations of nearest-neighbour bonds as a function of doping, the positive connected correlation at finite doping mostly originates from the presence of magnetic polarons formed by doped holes.

\begin{figure}
\includegraphics[]{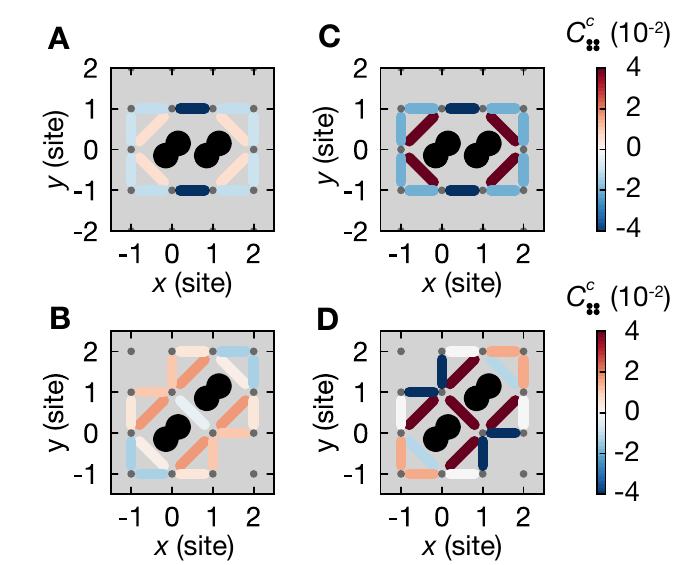}
\caption{\textbf{Influence of two doublon-hole fluctuations on the spin environment.} Connected spin correlations surrounding \textbf{A}, \textbf{C} nearest-neighbour and \textbf{B}, \textbf{D} diagonal doublon pairs, calculated from \textbf{A},\textbf{B} experimental data for $n \in [0.88,1.02]$ at $U/t=7.4(8)$ and \textbf{C}, \textbf{D} by exact diagonalization at half filling.}
\label{fig:figsi5}
\end{figure}

In a similar manner, we show connected spin correlations $C^{c}_{ \overset{\scalebox{0.45}{\newmoon}}{\scalebox{0.45}{\newmoon}}\overset{\scalebox{0.45}{\newmoon}}{\scalebox{0.45}{\newmoon}}  } $ surrounding nearest-neighbour and diagonal pairs of doublons close to half filling in Fig.~\ref{fig:figsi5} to characterize the effect of holes originating from doublon-hole fluctuations in the four-point correlations in the main text. There are two main observations relevant to our understanding of the four-point correlations presented in the main text Fig.~4. For nearest-neighbour doublons, a closest distance antiferromagnetic correlation is visible in the experiment and predicted by exact diagonalization (ED). This explains why the signal of Fig. 4C for the nearest-neighbour hole-pair shows an antiferromangetic signal at half filling. The second important insight concerns the connected correlation of the closest bond of diagonal doublon pairs, which is positive (ferromagnetic) in ED and shows a vanishing correlation value with experimental data. Therefore any connected antiferromagnetic correlation of this bond detected for diagonal hole pairs in Fig. 4 does not originate from doublon-hole contributions. Furthermore, close to half filling a positive signal from doublon-hole fluctuations might cancel a negative signal from doped holes and explain the uncorrelated value observed at very low doping in Fig. 4.


\subsection{Summary of experimental findings}
We summarize key experimental findings of the main manuscript in Table~S1. These phenomena lead us to the conclusion, that the onset of the Fermi liquid regime is $\delta_{\text{FL}}\sim30\,\%$. All stated dopings are broadly estimated values from the figures of the main manuscript. The crossover from polaronic metal to Fermi liquid cannot be assigned to one exact doping in our experiment.
\begin{table}[ht!]
\begin{tabular}{|l |l |l|}
	\hline
	\textbf{Observable} & \textbf{Doping} & \textbf{Behavior} \\ \hline
	spin-spin & $20$-$40\,\%$ & Various distances \\
	& & reverse sign \\ \hline
	$S(\boldsymbol{q})$ & $50\,\%$ & Visible incommensurate \\
	& & fluctuations \\ \hline
	$\chi_s$ & $20$-$30\,\%$ & Slope changes  \\ \hline
	hole-spin-spin & $20\,\%$ & correlation \\
	$(r=\sqrt{1.25},d=1)$ & & reverses sign \\ \hline
	hole-spin-spin & $20$-$40\,\%$ & correlation   \\
	$(r=\sqrt{0.5},\sqrt{2})$ & & maximally negative \\ \hline
	hole-hole-spin-spin & $30\,\%$ & correlation \\
	$(l=\sqrt{2},r=0,d=\sqrt{2})$ & & maximally negative \\ \hline
	hole-hole-spin-spin & $>30\,\%$ & agreement with \\
	& & free fermions (FL) \\ \hline
\end{tabular}
\caption{Summary of key experimental results.}
\end{table}

\subsection{Numerical calculations}
Connected correlations of uniform-RVB (uRVB), $\pi$-flux states and the string model or free fermions are computed from sampled snapshots with the same procedure as for experimental snapshots. The sampling procedure as well as ED, RPA and TPSC calculations are outlined below. The total number of snapshots used for [uRVB, $\pi$-flux, string, free] is [$4950$, $4600$, $5000$, $5000$] with system size $(L_x,L_y)=(16,16)$ sites.

\subsubsection{Exact diagonalization}

The exact diagonalization (ED) calculations compute the high-order correlation functions for the Hubbard model in a 4$\times$4 cluster with periodic boundary conditions. We keep only nearest-neighbor hopping $t$ and set $U=8t$ throughout the paper. To obtain the finite-temperature ($k_{B}T=0.4\,t$) results, we evaluate the expectation values of observables in a canonical ensemble, namely \begin{equation} \left\langle \hat{O}\right\rangle = \Tr\left[ \frac{e^{- \Ham/k_{B}T}}{\mathcal{Z}} \hat{O} \right] \approx \sum_{n<n_{\rm max}} \frac{e^{-E_n/k_{B}T}}{\mathcal{Z}}\langle n|\hat{O}|n\rangle, \end{equation} where the $\mathcal{Z}$ is the partition function. The $n_{\rm max}$ sets the numerical truncation of excited states, which satisfies $E_{n_{\rm max}}-E_0 \gg k_{B}T$. The excited states involve all total $S^z$ sectors. To determine these ground- and excited-state wavefunctions $|n\rangle$, we use the parallel Arnoldi method and the Paradeisos algorithm\cite{lehoucq1998arpack,jia2017paradeisos}.

The anomalous jump of the four-point correlation in the ED calculation (c.f. Fig.~4C) at 1/16 doping results from the finite-size effect when a single hole is doped into the 16-site cluster. It does not reflect the realistic correlator at a 6.26$\%$ doped thermodynamic system.

\subsubsection{Fermi liquid}
\emph{Free fermions.--} 
Theoretical predictions for non-interacting fermions can be obtained by applying Wick's theorem in the calculation of correlation functions. In an alternative to using Wick's theorem, which is closer to the experimental data, we produce snapshots in the Fock basis $\ket{\alpha}$ in the lattice. To this end we use Metropolis Monte-Carlo sampling on the distribution 
\begin{equation}
p_{\beta}(\alpha_{\vec{r}},\alpha_{\vec{k}}) = Z^{-1} e^{-\beta E(\alpha_{\vec{k}})} |\langle \alpha_{\vec{r}} | \alpha_{\vec{k}} \rangle |^2, 
\label{eqFreeFermionDistribution}
\end{equation}
where $| \alpha_{\vec{k}} \rangle$ are free-fermion wavefunctions, $e^{-\beta E(\alpha_{\vec{k}})}$ is the corresponding thermal weight and the overlaps $\langle \alpha_{\vec{r}} | \alpha_{\vec{k}} \rangle$ are Slater determinants which are easy to evaluate numerically. 

\emph{RPA.--} We go beyond free fermions by using the random phase approximation (RPA) \cite{Pines2018}, which allows us to calculate the spin- and charge susceptibilities, $\chi_{\rm s}(\vec{q},\omega)$ and $\chi_{\rm c}(\vec{q},\omega)$ respectively. Using the fluctuation-dissipation theorem with bosonic Matsubara frequencies $i \omega_m = i 2 \pi m / \beta$ (where $\beta = (k_B T)^{-1}$ and $m \in \mathbb{Z}$), the static structure factors $S(\vec{q})$ (spin) and $N(\vec{q})$ (charge) can be easily obtained:
\begin{flalign}
 S(\vec{q}) &= - \frac{T}{2} \sum_{i \omega_m} \chi_{\rm s}(\vec{q}, i \omega_m),\\
 N(\vec{q}) &= V n^2 \delta_{\vec{q},0} - 2 T \sum_{i \omega_m} \chi_{\rm c}(\vec{q}, i \omega_m),
\end{flalign}
where $V = L^2$ is the area of the system, $n = (N_\uparrow + N_\downarrow) / L^2$ is the total density and $\delta_{\vec{i},\vec{j}}$ the delta function. 

For free fermions, the spin- and charge- susceptibilities are equal, $\chi_{\rm s}(\vec{q},\omega) = \chi_{\rm c}(\vec{q},\omega) \equiv \chi_0(\vec{q},\omega)$, and given by the Lindhard function:
\begin{equation}
 \chi_0(\vec{q},i \omega_m) =  \sum_{\vec{p}} \frac{n_{\vec{p}}^{\rm F} -n_{\vec{p}+\vec{q}}^{\rm F} }{i \omega_m - \l \epsilon_{\vec{p} + \vec{q}} - \epsilon_{\vec{p} } \r},
\end{equation}
where $n_{\vec{p}}^{\rm F} = (1 + e^{\beta (\epsilon_{\vec{p}} - \mu)})^{-1}$ denotes the Fermi-Dirac distribution and $\epsilon_{\vec{p}} = - 2 t_x \cos(k_x) - 2 t_y \cos(k_y)$ is the free fermion dispersion relation in the lattice. 

For on-site Hubbard interactions $U$, the RPA expressions for the susceptibilities are given by \cite{Pines2018}
\begin{flalign}
 \chi_{\rm s}^{\rm RPA}(\vec{q},\omega) &= \frac{\chi_0(\vec{q},\omega)}{1 + \frac{U}{V} \chi_0(\vec{q},\omega)}, \\
 \chi_{\rm c}^{\rm RPA}(\vec{q},\omega) &= \frac{\chi_0(\vec{q},\omega)}{1 - \frac{U}{V} \chi_0(\vec{q},\omega)}.
\end{flalign}
Note that we used a convention where $\chi_0(\vec{q},\omega) \leq 0$; hence for sufficiently strong repulsive interactions $U>0$ and large enough densities $n$ the spin susceptibility diverges (Stoner instability). The charge susceptibility remains finite in this case. This divergence of the spin susceptibility is a result of neglecting renormalizations of the Hubbard interactions $U$ within the RPA. For RPA calculations in Fig. 2D of the manuscript we chose $U/t=4$, which matches the strongly doped experimental data ($\delta>40\,\%$) and diverges for intermediate dopings.

\emph{TPSC.--} We use the two-particle self-consistent (TPSC) way to include the renormalization of Hubbard interactions $U$ within the RPA formalism, following the proposal by Vilk et al. \cite{Vilk1994}, see also Ref.~\cite{Tremblay2011}. This approach assumes that the interaction vertices for spin and charge renormalize independently, which amounts to using different Hubbard $U$'s in the RPA expressions for the susceptibilities:
\begin{flalign}
 \chi_{\rm s}^{\rm TPSC}(\vec{q},\omega) &= \frac{\chi_0(\vec{q},\omega)}{1 + \frac{U_{\rm S}}{V} \chi_0(\vec{q},\omega)}, \\
 \chi_{\rm c}^{\rm TPSC}(\vec{q},\omega) &= \frac{\chi_0(\vec{q},\omega)}{1 - \frac{U_{\rm C}}{V} \chi_0(\vec{q},\omega)}.
\end{flalign}
For a given value of $U$ in the Hubbard model, the values of $U_{\rm S, C}$ are determined by demanding that the following local sum rules are satisfied,
\begin{flalign}
 - \frac{T}{V^2 n} \sum_{\vec{q}, i \omega_m} \l \chi_{\rm c}(\vec{q},i\omega_m) + \chi_{\rm s}(\vec{q},i\omega_m)  \r &= 2 n (1-n),  \label{eqLocSumRule1} \\
  - \frac{T}{V^2 n} \sum_{\vec{q}, i \omega_m} \l \chi_{\rm c}(\vec{q},i\omega_m) - \chi_{\rm s}(\vec{q},i\omega_m)  \r &= 4 n_{\uparrow \downarrow} - n^2, \label{eqLocSumRule2}
\end{flalign}
where $n_{\uparrow \downarrow} = \langle \hat{n}_{\vec{i}, \uparrow} \hat{n}_{\vec{i}, \downarrow} \rangle$. The local sum rules \eqref{eqLocSumRule1}, \eqref{eqLocSumRule2} reflect the Pauli principle and can be shown to be satisfied for the exact susceptibilities of the interacting model \cite{Tremblay2011}; they are violated by the RPA expressions, however. 

To solve Eqs.~\eqref{eqLocSumRule1}, \eqref{eqLocSumRule2} for $U_{\rm S}$ and $U_{\rm C}$, an expression for $n_{\uparrow \downarrow}$ is required. We follow \cite{Vilk1994,Tremblay2011} and make the ansatz
\begin{equation}
 n_{\uparrow \downarrow} = \frac{U_{\rm S}}{U} \langle \hat{n}_{\vec{i},\uparrow} \rangle \langle \hat{n}_{\vec{i},\downarrow} \rangle = \frac{1}{4}  \frac{U_{\rm S}}{U} n^2,
\end{equation}
where the last equation assumes spin balance, $N_\uparrow = N_\downarrow$, and translational invariance.

\subsubsection{Resonating valence bond states}
Shortly after the discovery of high-temperature superconductivity in the cuprate materials, Anderson proposed the resonating valence bond (RVB) states as a possible description of these systems \cite{Anderson1987}. We simulate such RVB states by sampling Fock space snapshots from the Gutzwiller projected thermal density matrix of the mean-field Hamiltonian
\begin{equation}
\begin{split}
&\hat{\mathcal{H}}_{\rm MF} =
\\
 &-\frac{1}{2} t^* \sum_{\vec{i} \in A} \sum_\sigma \left( e^{i\theta_0} \hat{c}_{\vec{i},\sigma}^\dagger \hat{c}_{\vec{i}+\vec{x},\sigma} + e^{-i\theta_0} \hat{c}_{\vec{i},\sigma}^\dagger \hat{c}_{\vec{i}+\vec{y},\sigma} + h.c. \right) 
\\
&-\frac{1}{2} t^* \sum_{\vec{i} \in B} \sum_\sigma \left( e^{-i\theta_0} \hat{c}_{\vec{i},\sigma}^\dagger \hat{c}_{\vec{i}+\vec{x},\sigma} + e^{i\theta_0} \hat{c}_{\vec{i},\sigma}^\dagger \hat{c}_{\vec{i}+\vec{y},\sigma} + h.c. \right).
\label{eqHMF}
\end{split}
\end{equation}
Here, $\vec{i} \in A(B)$ denotes lattice sites $\vec{i}$ which are part of the A(B) sublattice and $\hat{c}_{\vec{i},\sigma}^{(\dagger)}$ is the annihilation (creation) operator of a fermion with spin $\sigma$. The mean-field Hamiltonian describes a system with staggered flux $\pm \Phi = \pm 4 \theta_0$ and effective hopping amplitude $t^*$. In particular, we consider uniform RVB states, for which $\theta_0 = 0$, and $\pi$-flux RVB states with $\theta_0=\pi/4$.

In order to obtain real space snapshots, we simultaneously sample real space configurations $|\tilde{\alpha}_{\vec{r}}\rangle$ and momentum space configurations $| \alpha_{\vec{k}} \rangle$. In momentum space, the two spin species are treated separately, such that two fermions of opposite spin can occupy the same momentum state. In real space, we directly apply the Gutzwiller projection during sampling: each site can only be empty or occupied with a spin up or a spin down fermion. Since the mean field Hamiltonian \eqref{eqHMF} can be readily diagonalized in momentum space, we obtain an energy $E(\alpha_{\vec{k}})$ for each $\vec{k}$-space configuration and thus the corresponding thermal weight. We use the Metropolis Monte Carlo algorithm \cite{Gros1989} to sample Gutzwiller projected real space snapshots $| \tilde{\alpha}_{\vec{r}} \rangle$ according to the probability distribution
\begin{equation}
p_{\beta}(\tilde{\alpha}_{\vec{r}},\alpha_{\vec{k}}) = Z^{-1} e^{-\beta E(\alpha_{\vec{k}})} | \langle \tilde{\alpha}_{\vec{r}} | \alpha_{\vec{k}} \rangle |^2.
\end{equation}
The temperature is set to $k_{B}T=0.4\,t^{\star}$. This procedure is identical as for free fermions, see Eq.~\eqref{eqFreeFermionDistribution}, except for the fact that we constrain ourselves to Fock states $| \tilde{\alpha}_{\vec{r}} \rangle$ with maximally one fermion per site.

\subsubsection{Geometric string theory}
In the geometric string theory picture, we assume all dopants to be magnetic polarons that do not interact with each other. A single dopant is described using the geometric string theory, which is based on a Born-Oppenheimer-type approximation: the Hilbert space is approximated as a tensor product of the spinon and chargon Hilbert space \cite{Grusdt2018,Chiu2019,Koepsell2019}. The Hamiltonian in this effective Hilbert space is then given by the kinetic energies (hopping) of the spinon and chargon, as well as a linear string potential confining the spinon to the chargon. The corresponding linear string tension is determined from nearest, straight and diagonal next-nearest spin correlations in the undoped system \cite{Grusdt2018}. 

The resulting spinon-chargon problem can be readily solved and thus a string length distribution is obtained, where the string length is the number of bonds the chargon moves on top of the unperturbed spin background. For each doping value, we start from a set of 5000 quantum Monte Carlo snapshots of the Heisenberg model at $T/J=0.8$ and put in the corresponding number of holes by hand. For each hole, we sample a string length from the thermal distribution and move the hole for the corresponding number of bonds. This procedure was previously described in \cite{Chiu2019}.

\subsubsection{DMRG simulations of the $t-J$ model}

We calculate the ground state of the $t-J$ model for $t/J=2$ on a $6 \times 18$ cylinder with periodic boundary conditions in the short direction using the TeNPy package \cite{hauschildTenpy,Hauschild2018SciPost}. We use particle and $S^{z,\rm tot}$ conservation and work in the sector with $S^{z,\rm tot}=0$ and two holes.


\bigskip

\begin{thebibliography}{10}

\bibitem{Keimer2015}
B.~Keimer, S.~A. Kivelson, M.~R. Norman, S.~Uchida, J.~Zaanen, {\it Nature\/}
  {\bf 518}, 179 (2015).

\bibitem{Dagotto1994}
E.~Dagotto, {\it Reviews of Modern Physics\/} {\bf 66}, 763 (1994).

\bibitem{Lee2006}
P.~A. Lee, N.~Nagaosa, X.-G. Wen, {\it Reviews of Modern Physics\/} {\bf 78},
  17 (2006).

\bibitem{Badoux2016}
S.~Badoux, {\it et~al.\/}, {\it Nature\/} {\bf 531}, 210 (2016).

\bibitem{Chen2019}
S.~D. Chen, {\it et~al.\/}, {\it Science\/} {\bf 366}, 1099 (2019).

\bibitem{Doiron2007}
N.~Doiron-Leyraud, {\it et~al.\/}, {\it Nature\/} {\bf 447}, 565 (2007).

\bibitem{Yang2009}
H.~B. Yang, {\it et~al.\/}, {\it Physical Review Letters\/} {\bf 107} (2011).

\bibitem{Ronning2005}
F.~Ronning, {\it et~al.\/}, {\it Physical Review B\/} {\bf 71} (2005).

\bibitem{Schrieffer2007}
J.~R. Schrieffer, {\it {Handbook of High-Temperature Superconductivity}\/}
  (Springer, New York, 2007).

\bibitem{Bulaevski1968}
L.~Bulaevski, {\'{E}}.~Nagaev, D.~Khomskiǐ, {\it Soviet Journal of
  Experimental and Theoretical Physics\/} {\bf 27}, 836 (1968).

\bibitem{SchmittRink1988}
S.~Schmitt-Rink, C.~M. Varma, A.~E. Ruckenstein, {\it Physical Review
  Letters\/} {\bf 60}, 2793 (1988).

\bibitem{Shraiman1988}
B.~I. Shraiman, E.~D. Siggia, {\it Physical Review Letters\/} {\bf 61}, 467
  (1988).

\bibitem{Kane1989}
C.~L. Kane, P.~A. Lee, N.~Read, {\it Physical Review B\/} {\bf 39}, 6880
  (1989).

\bibitem{Sachdev1989}
S.~Sachdev, {\it Phys. Rev. B\/} {\bf 39}, 12232 (1989).

\bibitem{Grusdt2018}
F.~Grusdt, {\it et~al.\/}, {\it Phys. Rev. X\/} {\bf 8}, 11046 (2018).

\bibitem{Blomquist2019}
E.~Blomquist, J.~Carlstr{\"{o}}m, {\it arXiv:1912.08825\/}  (2019).

\bibitem{Frachet2020}
M.~Frachet, {\it et~al.\/}, {\it Nature Physics\/}  (2020).

\bibitem{LeBlanc2015}
P.~F. LeBlanc, {\it et~al.\/}, {\it Physical Review X\/} {\bf 5} (2015).

\bibitem{Chen2020}
B.-B. Chen, {\it et~al.\/}, {\it arXiv.2008.02179\/}  (2020).

\bibitem{Gross2017}
C.~Gross, I.~Bloch, {\it Science\/} {\bf 357}, 995 (2017).

\bibitem{Nichols2018}
M.~A. Nichols, {\it et~al.\/}, {\it Science\/} {\bf 363}, 383 (2019).

\bibitem{Brown2019a}
P.~T. Brown, {\it et~al.\/}, {\it Science\/} {\bf 363}, 379 (2019).

\bibitem{Ji2020}
G.~Ji, {\it et~al.\/}, {\it arXiv:2006.06672\/}  (2020).

\bibitem{Chiu2019}
C.~S. Chiu, {\it et~al.\/}, {\it Science\/} {\bf 365}, 251 (2019).

\bibitem{Hartke2020}
T.~Hartke, B.~Oreg, N.~Jia, M.~Zwierlein, {\it arXiv:2003.11669\/}  (2020).

\bibitem{Mazurenko2017}
A.~Mazurenko, {\it et~al.\/}, {\it Nature\/} {\bf 545}, 462 (2017).

\bibitem{Boll2016}
M.~Boll, {\it et~al.\/}, {\it Science\/} {\bf 353}, 1257 (2016).

\bibitem{Koepsell2020}
J.~Koepsell, {\it et~al.\/}, {\it Physical Review Letters\/} {\bf 125}, 10403
  (2020).

\bibitem{Koepsell2019}
J.~Koepsell, {\it et~al.\/}, {\it Nature\/} {\bf 572}, 358 (2019).

\bibitem{Vijayan2019}
J.~Vijayan, {\it et~al.\/}, {\it Science\/} {\bf 367}, 186 (2019).

\bibitem{Salomon2019}
G.~Salomon, {\it et~al.\/}, {\it Nature\/} {\bf 565}, 56 (2019).

\bibitem{Hilker2017}
T.~A. Hilker, {\it et~al.\/}, {\it Science\/} {\bf 357}, 484 (2017).

\bibitem{SM}
{\it {see Supplementary Material}\/}.

\bibitem{Schweigler2017}
T.~Schweigler, {\it et~al.\/}, {\it Nature\/} {\bf 545}, 323 (2017).

\bibitem{Anderson1987}
P.~W. Anderson, {\it Science\/} {\bf 235}, 1196 (1987).

\bibitem{Vilk1994}
Y.~M. Vilk, L.~Chen, A.~M. Tremblay, {\it Physical Review B\/} {\bf 49}, 13267
  (1994).

\bibitem{Moreo1990}
A.~Moreo, D.~J. Scalapino, R.~L. Sugar, S.~R. White, N.~E. Bickers, {\it
  Physical Review B\/} {\bf 41} (1990).

\bibitem{Furukawa1992}
N.~Furukawa, M.~Imada, {\it Journal of the Physical Society of Japan\/} {\bf
  61}, 3331 (1992).

\bibitem{Cheong1991}
S.~W. Cheong, {\it et~al.\/}, {\it Physical Review Letters\/} {\bf 67}, 1791
  (1991).

\bibitem{Drewes2016}
J.~H. Drewes, {\it et~al.\/}, {\it Physical Review Letters\/} {\bf 117} (2016).

\bibitem{Moreo1993}
A.~Moreo, {\it Physical Review B\/} {\bf 48} (1993).

\bibitem{Preuss1997}
R.~Preuss, W.~Hanke, C.~Gr{\"{o}}ber, H.~G. Evertz, {\it Physical Review
  Letters\/} {\bf 79}, 1122 (1997).

\bibitem{Schrieffer1989}
J.~R. Schrieffer, X.~. Wen, S.~C. Zhang, {\it Physical Review B\/} {\bf 39}, 11
  663 (1989).

\bibitem{Blomquist2020}
E.~Blomquist, J.~Carlstr{\"{o}}m, {\it arXiv:2007.15011\/}  (2020).

\bibitem{White1997a}
S.~R. White, D.~Scalapino, {\it Physical Review B\/} {\bf 55}, R14701 (1997).

\bibitem{Kantian2016}
A.~Kantian, S.~Langer, A.~J. Daley, {\it Physical Review Letters\/} {\bf 120},
  060401 (2018).

\bibitem{Khatami2011}
E.~Khatami, M.~Rigol, {\it Physical Review A\/} {\bf 84}, 53611 (2011).

\bibitem{Bohrdt2020}
A.~Bohrdt, {\it et~al.\/}, {\it arXiv:2007.07249\/}  (2020).

\bibitem{Punk2015}
M.~Punk, A.~Allais, S.~Sachdev, {\it Proceedings of the National Academy of
  Sciences\/} {\bf 112}, 9552 (2015).

\bibitem{Zhang2020}
Y.-H. Zhang, S.~Sachdev, {\it arXiv.2001.09159\/}  (2020).

\newcounter{enumi_saved}
\setcounter{enumi_saved}{\value{NAT@ctr}}

\end{thebibliography}

\begin{thebibliography}{10}
\setcounter{NAT@ctr}{\value{enumi_saved}}

\bibitem{Omran2015}
A.~Omran, {\it et~al.\/}, {\it Physical Review Letters\/} {\bf 115}, 1 (2015).

\bibitem{ColemanBook}
P.~Coleman, {\it {Introduction to Many-Body Physics}\/} (Cambridge University
  Press, Cambridge, 2015).

\bibitem{Zhou2011}
Q.~Zhou, T.~L. Ho, {\it Physical Review Letters\/} {\bf 106} (2011).

\bibitem{lehoucq1998arpack}
R.~B. Lehoucq, D.~C. Sorensen, C.~Yang, {\it {ARPACK Users' Guide: Solution of
  Large-Scale Eigenvalue Problems with Implicitly Restarted Arnoldi Methods}\/}
  (Siam, 1998).

\bibitem{jia2017paradeisos}
C.~J. Jia, Y.~Wang, C.~B. Mendl, B.~Moritz, T.~P. Devereaux, {\it Comput. Phys.
  Commun.\/} {\bf 224}, 81 (2018).

\bibitem{Pines2018}
D.~Pines, {\it {Theory of Quantum Liquids: Normal Fermi Liquids}\/} (CRC Press,
  2018).

\bibitem{Tremblay2011}
A.-M.~S. Tremblay, {\it Strongly Correlated Systems: Theoretical Methods\/},
  A.~Avella, F.~Mancini, eds. (Springer Berlin Heidelberg, Berlin, Heidelberg,
  2012), pp. 409--453.

\bibitem{Gros1989}
C.~Gros, {\it Annals of Physics\/} {\bf 189}, 53 (1989).

\bibitem{hauschildTenpy}
J.~Hauschild, {\it et~al.\/}, {\it {Tensor network python}\/} (The code is
  available online at https://github.com/tenpy/tenpy/, the documentation can be
  found at https://tenpy.github.com/, 2018).

\bibitem{Hauschild2018SciPost}
J.~Hauschild, F.~Pollmann, {\it SciPost Physics Lecture Notes\/} {\bf 5}, 005
  (2018).

\end{thebibliography}

\end{document}